\documentclass[12pt,preprint]{aastex}

\newcommand{\msun} {$M_{\sun}$}

\newcommand{\Te} {T_{\rm eff}}
\newcommand{\logg} {\log g}

\begin{document}


\title{NEW INSIGHTS INTO THE PROBLEM OF THE SURFACE GRAVITY DISTRIBUTION OF COOL DA WHITE DWARFS}

\author{P.-E. Tremblay$^{1}$, P. Bergeron$^{1}$, J.~S. Kalirai$^{2}$, and A. Gianninas$^{1}$}
\affil{$^{1}$D\'epartement de Physique, Universit\'e de Montr\'eal, C.P.~6128, 
Succ.~Centre-Ville, Montr\'eal, QC H3C 3J7, Canada}
\affil{$^{2}$Space Telescope Science Institute, 3700 San Martin Drive, Baltimore, MD 21218, USA}
\email{tremblay@astro.umontreal.ca, bergeron@astro.umontreal.ca, jkalirai@stsci.edu, gianninas@astro.umontreal.ca}

\begin{abstract}

We review at length the longstanding problem in the spectroscopic
analysis of cool hydrogen-line (DA) white dwarfs ($\Te < 13,000$~K)
where gravities are significantly higher than those found in hotter DA
stars. The first solution that has been proposed for this problem is a
mild and systematic helium contamination from convective mixing that
would mimic the high gravities. We constrain this scenario by
determining the helium abundances in six cool DA white dwarfs using
high-resolution spectra from the Keck I 10-m telescope. We obtain no
detections, with upper limits as low as He/H = 0.04 in some
cases. This allows us to put this scenario to rest for good. We also
extend our model grid to lower temperatures using improved Stark
profiles with non-ideal gas effects from Tremblay \& Bergeron and find
that the gravity distribution of cool objects remains suspiciously
high. Finally, we find that photometric masses are, on average, in
agreement with expected values, and that the high-$\logg$ problem is
so far unique to the spectroscopic approach.

\end{abstract}


\section{ASTROPHYSICAL CONTEXT}

The most accurate method for determining the atmospheric parameters of
hydrogen-line (DA) white dwarfs --- $\Te$ and $\log g$ --- is through
a detailed comparison of the observed Balmer line profiles with the
predictions of model atmospheres. This so-called spectroscopic
technique was applied to a sample of 37 cool ($\Te\lesssim 12,000$~K)
DA stars by \citet{bergeron90} who showed that the surface gravities
inferred from spectroscopic measurements were significantly larger
than the canonical value of $\logg\sim8$ expected for these
stars. This result was readily interpreted as evidence for convective
mixing between the thin superficial hydrogen layer with the more
massive underlying helium envelope, a process originally, and
independently, proposed by \citet{koester76}, \citet{vauclair77}, and
\citet{dantona79}. In this mixing process, helium brought to the
surface would remain spectroscopically invisible because of the low
photospheric temperatures. However, the helium abundance can still be
determined from a detailed examination of the high Balmer lines, since
the presence of helium increases the photospheric pressure, and thus
produces a quenching of the upper levels of the hydrogen atom which,
in turn, affects the line profiles
\citep{liebert83}. \citet{bergeron91} showed on a more quantitative
basis that the effects produced on the hydrogen lines at high $\log g$
--- or high masses --- could not be distinguished from those produced
by the presence of large amounts of helium in the atmospheres of cool
DA stars. Such helium-enriched DA white dwarfs would simply {\it
appear} to have high surface gravities when analyzed under the
assumption of pure hydrogen atmospheres. By assuming instead a
canonical value of $\logg=8$ for all stars in their sample,
\citet{bergeron90} were able to demonstrate that the surface
compositions of most cool DA white dwarfs are contaminated by large
amounts of helium, sometimes as high as He/H$~\sim20$.

This astrophysical interpretation, however, rests heavily on the
theoretical framework, based on the occupation probability formalism
of \citet{hm88}, which was implemented by \citet{bergeron91} to
properly model the hydrogen line profiles. It is thus entirely
plausible that these improved model spectra yield large spectroscopic
$\logg$ values simply because of inaccuracies in the input
physics. With this idea in mind, \citet[][hereafter BSL92]{bergeron92}
applied the spectroscopic technique to a large sample of DA white
dwarfs that are sufficiently hot ($\Te\gtrsim 13,000$~K) to avoid any
theoretical uncertainties related to the onset of convective energy
transport at low effective temperatures, and to the possible presence
of spectroscopically invisible traces of helium. The assumptions of
radiative atmospheres with pure hydrogen compositions were then
perfectly justified. The spectroscopic analysis of BSL92 allowed the
first measurement of the $\logg$ and mass distributions of DA stars
with an unprecedented accuracy. In particular, the spectroscopic
$\logg$ distribution showed a sharp peak with a mean value of 7.909
with a dispersion of only 0.257, in agreement with the expected mean
value for DA stars. Hence the spectroscopic technique and model
spectra seemed to produce reasonably sound results, at least in this
temperature range.

Around the same time, \citet{bwf92} showed that the assumed
parametrization of the convective efficiency, treated within the
mixing-length theory in white dwarf models, could affect the emergent
fluxes, and in particular the predicted line profiles of DA stars in
the range $\Te\sim8000 - 14,000$~K. This sensitivity suggests that
perhaps the earlier interpretation of the presence of invisible traces
of helium in cool DA stars could be caused by an inadequate
parametrization of the mixing-length theory. In an attempt to properly
calibrate the convective efficiency in the atmospheres of DA stars,
\citet[][hereafter B95]{bergeron95} studied a sample of 22 ZZ Ceti
stars and showed that the so-called ML2/$\alpha=0.6$ version of the
mixing-length theory (where $\alpha$ is the mixing-length to pressure
scale height ratio) yielded the best internal consistency between
optical and UV effective temperatures, trigonometric parallaxes, $V$
magnitudes, and gravitational redshift
measurements\footnote{\citet{koester94} arrived at a similar
conclusion with a somewhat equivalent version of the mixing-length
theory, ML1/$\alpha=2.0$, although their analysis was based on a
single object.}. Yet, despite this adopted calibration, the large
$\logg$ values observed in cool DA stars remained.

Our current view of what we will now refer to as the ``high-$\logg$
problem'' is best summarized in Figure \ref{fg:f1}, where we show the
distribution of surface gravity as a function of effective temperature
for the sample of over 1200 relatively bright DA white dwarfs of
\citet{gianninas09}, drawn from the online version of the Villanova
White Dwarf
Catalog\footnote{http://www.astronomy.villanova.edu/WDCatalog/index.html}
\citep[see also similar results
in][]{liebert05,bergeron07,kepler07,koester09,gianninas09}. The model
spectra used to determine our atmospheric parameters rely on the
improved Stark broadening profiles of \citet[][hereafter
TB09]{tremblay09}; these models are discussed further in \S~3. While
the hotter DA stars in Figure \ref{fg:f1} follow a distribution close
to the canonical mean of $\logg=8$, the surface gravities are shown to
increase suddenly below $\Te\sim12,500$~K to a constant average value
of $\logg\sim8.2$. Such high $\logg$ values cannot be real, of course,
since white dwarf stars are expected to cool at constant
radii. Furthermore, DA white dwarfs at $\Te \sim 12,500$~K have
cooling ages of the order of 1 Gyr or so, and they can hardly
represent a distinct galactic population (see additional discussions
in \citealt{kepler07} and \citealt{koester09}). Consequently,
inaccuracies in the model atmosphere calculations or inadequate
assumptions about the composition of these stars need to be called
upon to account for the observed trend. This longstanding problem
represents a serious hurdle to obtaining reliable atmospheric
parameters at the cool end of the white dwarf evolutionary
sequence. This, in turn, may affect our determination of the white
dwarf luminosity function \citep[e.g.,][]{harris06} and our ability to
use white dwarfs as reliable cosmochronometers and distance indicators
\citep{fontaine01,hansen07}.

\citet{bergeron07} and \citet{koester09} have produced the most recent
and extensive reviews of the high-$\logg$ problem and the possible
solutions, but none appear very satisfactory to date. In particular,
Koester concludes that inaccuracies in the treatment of convection in
model atmospheres is responsible for the problem, for the lack of a
better alternative. The mixing-length theory currently used in the
models is indeed a crude approximation that makes use of a free
parameter, but model spectra properly calibrated yield atmospheric
parameters that are quite reasonable, in particular for stars near the
ZZ Ceti instability strip \citep[see, e.g.,][]{gianninas06}.
Furthermore, \citet{ludwig94} showed with more sophisticated 2D
hydrodynamic models that the mixing-length theory was roughly correct
to predict the temperature structure of the photospheric regions of DA
white dwarfs. More detailed comparisons may prove otherwise,
however. Their results do suggest that deeper layers have a
significantly higher convective efficiency than predicted from the
mixing-length theory, but this has little effect on the predicted
spectra.

The solution originally proposed by \citet{bergeron90} that the
presence of helium in the atmospheres of cool DA stars is responsible
for the high-$\logg$ problem cannot be tested directly since helium
becomes spectroscopically invisible at the effective temperatures and
expected helium-to-hydrogen abundance ratios where the problem
manifests itself. Or at least, this used to be our common
understanding. This view has recently been challenged by two important
discoveries. \citet{koester05} were the first to report the detection
of weak helium lines in a DA star (HS 0146+1847) with a temperature
slightly below $T_{\rm eff}\sim13,000$~K; this white dwarf turns out
to be a peculiar helium-dominated and metal-rich DAZB star. More
importantly, HIRES spectroscopic observations with Keck of
the metal-rich DAZ star GD~362 revealed the presence of a very weak
He~\textsc{i} $\lambda5877$ absorption feature \citep{zuckerman07}, in
a white dwarf with an estimated temperature of only
$\Te\sim10,000$~K. The detailed model atmosphere analysis by Zuckerman
et al.~also revealed that GD 362 has in fact a helium-dominated
atmosphere. Interestingly enough, GD 362 and HS 0146+1847 are also
surrounded by a circumstellar disk \citep{becklin05,farihi09}. These
findings clearly demonstrate that helium can indeed be detected {\it
directly} in this temperature range when observed at sufficiently
high signal-to-noise ratio and high dispersion. The weak helium lines
detected in both white dwarfs could even be reproduced with model
spectra.

In this paper, we provide new insights into the problem of the surface
gravity distribution of cool DA white dwarfs by presenting
observations of six objects with the High Resolution Echelle
Spectrometer \citep[HIRES,][]{vogt94} on the Keck I 10-m telescope and
attempt to measure, or constrain, {\it directly} the helium abundance
in the atmosphere of DA white dwarfs.

\section{THE PRESENCE OF HELIUM IN COOL DA WHITE DWARFS}

\subsection{A Reassessment of the Convective Mixing Scenario}

As discussed above, the contamination of DA white dwarf atmospheres by
trace amounts of helium has been the first and longest running
solution to the high-$\logg$ problem. In particular,
\citet{bergeron91} showed that at low effective temperatures
($\Te\lesssim 12,000$~K), a high surface gravity DA white dwarf with a
pure hydrogen atmosphere would appear identical to a helium-enriched
DA star with a normal mass. On the right panels of Figure \ref{fg:f2},
we illustrate a typical example of this degeneracy for the cool DA
star LHS 3254, where the spectroscopic fit with pure hydrogen models
($\Te=9370$~K, $\log g=8.27$) is qualitatively equivalent to the fit
with a mixed helium/hydrogen composition of ${\rm He/H}=1.0$ at a
comparable temperature ($\Te=9300$~K), but with a much lower $\logg$
value of 7.81. On the left panels are shown the corresponding fits to
the optical $BVRI$ and infrared $JHK$ photometry. Here, only the
effective temperature and the solid angle $\pi(R/D)^2$ are considered
free parameters ($R$ is the radius of the star and $D$ its distance
from Earth), while the distance is obtained from the trigonometric
parallax measurement. It can be seen that the $\Te$ and $\logg$ values
obtained from the photometric solutions are not as sensitive to the
assumed helium abundance as the spectroscopic solutions. In the
example shown here, the helium abundance has in fact been {\it
adjusted} to ensure that the photometric and spectroscopic $\logg$
values agree. We also note that the photometric and spectroscopic
temperatures agree even better when helium is included. Hence,
photometric and spectroscopic measurements combined with trigonometric
parallaxes can effectively provide, in principle, a unique and
self-consistent solution for the atmospheric parameters of cool DA
stars, assuming of course that the presence of helium is responsible
for the high-$\logg$ problem. Unfortunately, parallax measurements are
generally not accurate enough, or even unavailable, and there are many
instances where photometric $\logg$ values are already larger than
spectroscopic values, even with pure hydrogen models (see
\citealt{boudreault05} and the paragraph below).

With this in mind, it is generally preferable to use a statistical
approach to quantify the He/H ratio in cool DA stars. If we assume for
instance that the masses of cool DA stars are normal --- i.e. the same
as for hot DA stars --- we can fix the mass and consider the effective
temperature and the He/H ratio as free parameters in the spectroscopic
fitting technique. The results of this experiment are presented in
Figure \ref{fg:f3}, where we show the helium abundance as a function
of $\Te$ for the same sample of cool DA stars used in Figure
\ref{fg:f1}, but with the mass of each star set to 0.649~\msun, which
corresponds to the mean mass derived by TB09 for the DA stars above
$\Te = 13,000$~K in the Palomar-Green sample. The upper limits at the
bottom of the figure correspond to objects for which the spectroscopic
masses are already below the mean value adopted here. We can clearly
see that there is a concentration of stars with helium abundances in
the range He/H $\sim 0.1 - 0.3$. These results suggest that a mild but
{\it systematic} helium contamination could shift the mass
distribution to the expected lower value. For a better illustration,
we refitted the same sample of cool DA stars with a constant value of
He/H = 0.25 --- the average helium abundance in this sample --- and
plotted the results in Figure \ref{fg:f4} together with the pure
hydrogen solutions. Hence, with a sudden onset of helium contamination
near $\Te\sim 12,500$~K, we can obtain a stable mass distribution
throughout the entire white dwarf temperature regime. It must be
stressed that this systematic contamination does not prevent {\it
individual} objects like GD~362 from having much larger helium
abundances, or even some stars from having pure hydrogen
atmospheres. These objects must be rare, however, since the constant
mass distribution achieved in Figure \ref{fg:f4} can only be recovered
if the vast majority of cool DA stars have a mild helium
contamination.

As discussed above, the source of this helium contamination could be
most easily explained by convective mixing between the thin hydrogen
layer and the deeper and more massive helium envelope. The extent of
the hydrogen convection zone is displayed in Figure \ref{fg:f5} as a
function of effective temperature for a 0.6 \msun\ DA white dwarf,
based on evolutionary models with thick hydrogen layers similar to
those described by \citet{fontaine01}. These calculations show that if
the hydrogen envelope is thin enough, the bottom of the hydrogen
convection zone may eventually reach the underlying and more massive
convective helium layer, resulting in a mixing of the hydrogen and
helium layers. This figure also indicates that the effective
temperature at which mixing occurs will depend on the thickness of the
hydrogen envelope. The thicker the envelope, the lower the mixing
temperature; if the hydrogen layer is more massive than $M_{\rm
H}/M_{\rm tot} \sim 10^{-6}$, mixing will never occur. The simplest
physical model that can be used to describe this convective mixing
process is to assume that hydrogen and helium are homogeneously
mixed. Since the helium convection zone is much more massive ($M_{\rm
He-conv}/M_{\rm tot}\sim10^{-6}$) than the hydrogen layer when
mixing occurs, it is generally assumed that a DA star would be
transformed into a helium-atmosphere non-DA white dwarf with only a
trace abundance of hydrogen. By performing a statistical analysis of
the ratio of hydrogen- versus helium-rich white dwarfs as a function
of $\Te$, \citet{tremblay08} demonstrated that such mixing does indeed
occur, but not until white dwarfs reach temperatures below $\sim
9000$~K. Even then, mixing seems to occur for only 15\% of all DA
stars. This view is consistent with our current understanding that
most DA stars probably have thick hydrogen layers with $M_{\rm
H}/M_{\rm tot}\gtrsim 10^{-6}$ \citep[see Table 7 of][]{fb08}.

Since the helium abundances inferred from Figure \ref{fg:f3} are
significantly lower than those expected from this standard convective
mixing scenario, it is necessary to invoke an alternative and
incomplete mixing process that would systematically contaminate the
atmospheres of DA stars with only small amounts of helium, perhaps
through overshooting. However, the hydrogen convection zone in a
12,000~K DA white dwarf is very thin according to Figure \ref{fg:f5},
much smaller than the $\sim 10^{-6}$ \msun\ or so expected for the
total hydrogen layer mass of most DA stars, and any type of convective
overshooting thus appears improbable. If we were to confirm such an
incomplete mixing scenario, we would have to review our assumption
that hydrogen and helium are homogeneously mixed when convection zones
connect, as well as our estimates of the thickness of the hydrogen
layers in DA stars.

\subsection{Direct Detection of Helium in Cool DA White Dwarfs}

According to the incomplete convective mixing scenario, a systematic
helium contamination estimated at He/H = 0.25 (see Fig.~\ref{fg:f3})
would be able to explain the high-$\logg$ problem. Obviously, the
actual ratio may vary from star to star, but on average, the helium
abundance of individual DA stars must be close to this value if this
interpretation is correct. In this section, we attempt to test this
scenario through direct observations of helium in the photospheres of
cool DA stars. For this purpose, we used our model atmospheres to
predict the strength of the He~\textsc{i} $\lambda5877$ line for a
wide range of atmospheric parameters. These are similar to those
described in TB09, with the exception that we now include the detailed
helium line opacity calculations of \citet[][and references
therein]{beauchamp96} and \citet{beauchamp97}, which also take into
account the Hummer-Mihalas occupation probability formalism (more
details are provided in \S~3). In line with the results for GD 362 and
HS~0146+1847, we find that helium can indeed be detected
spectroscopically in the range of temperatures where the incomplete
mixing scenario is believed to occur, but the detection of this helium
line requires high-resolution ($R\sim 20,000$) and high
signal-to-noise ($S/N \gtrsim 100$) spectroscopic observations. Since
the brightest white dwarfs in this temperature range have $V \sim 12.5
- 15.0$, an echelle spectrograph on an 8-meter class telescope is
absolutely necessary to detect this weak absorption feature.

With these predictions in hand, we requested one night of observing
time on the Keck I 10-m telescope on Mauna Kea with HIRES. Our
objective was to observe the spectral region around the He~\textsc{i}
$\lambda5877$ line for at least five DA stars with $\Te < 12,500$~K to
investigate whether the helium contamination is indeed systematic. We
first computed the $S/N$ and integration time required to achieve a
3$\sigma$ detection for DA stars selected from the sample of
\citet{gianninas09}. We found that stars below $\Te \sim 10,500$~K
would require more than five hours of observation time to achieve our
detection threshold, even for the brightest stars, compromising our
original strategy. We thus reduced our target list to a representative
sample of six bright DA stars in the range $12,500 \gtrsim \Te \gtrsim
10,500$~K. The atmospheric parameters for all six white dwarfs are
given in Table 1; these are obtained by fitting lower resolution
spectra taken from the spectroscopic survey of \citet{gianninas09}
with pure hydrogen model atmospheres. The location of these objects in
Figure \ref{fg:f1} (circles) indicates that they occupy a region
in $\Te$ and $\log g$ where the spectroscopic $\logg$ values rise
significantly. We note that four DA stars in our sample have surface
gravities significantly in excess of the canonical value of
$\logg=8$. Our coolest target, G67-23 (labeled in Fig.~\ref{fg:f3}),
is predicted to be a very massive ($M\sim 1.14$ \msun) white dwarf
under the assumption of a pure hydrogen composition. Not surprisingly,
five out of six objects also turn out to be ZZ Ceti pulsators.

High $S/N$ spectroscopic observations for all stars in Table 1 were
secured on 2009, August 26 using the HIRES instrument with the red
collimator and the 0."86 slit ($R\sim50,000$). The spectroscopic data
were reduced using the MAKEE reduction software (version 5.2.4) and
carefully calibrated to vacuum wavelengths. We show in Figure
\ref{fg:f6} the high-resolution spectra for the six stars in the
region of the He~\textsc{i} $\lambda5877$ line, along with the
predicted helium line profiles. The thick solid lines correspond to
the predictions with a systematic helium contamination of He/H = 0.25,
while the dotted lines assume a 0.649 \msun\ stellar mass. This last
assumption implies that any surface gravity measurement above
$\logg\sim8.08$ is attributed to the presence of helium; this applies
to only four stars in Table 1 and in Figure \ref{fg:f6}. {\it Our
results clearly indicate a non-detection of helium for all objects
in our sample}.

Our spectroscopic data can provide upper limits on the helium
abundance in the photosphere of these stars. We first obtain from our
low-resolution spectra the best fitted parameters $\Te$ and $\logg$
for a given helium abundance. We then increase the helium abundance
until the depth of the predicted He~\textsc{i} $\lambda5877$ line
profile is 3$\sigma$ above the noise level of our HIRES
data\footnote{Since we oversample the predicted line profiles with the
0.1 \AA~resolution, we binned by a factor of 4 to provide better,
but still conservative upper limits.}. These upper limits on the
helium abundance are reported in Table 1 for each star. For five
objects, we find very low upper limits for the He/H abundance ratio
(less than one tenth). We must therefore conclude that a mild and
systematic helium contamination in the atmospheres of cool DA white
dwarfs cannot be the origin of the high-$\logg$ problem.

The case of G67-23 (2246+223) is particularly of interest since our
low-resolution optical spectrum exhibits high Balmer lines
characteristics of a massive DA star \citep[see Fig.~1b
of][]{lajoie07}, or alternatively, of a normal mass white dwarf with a
helium-dominated atmosphere, such as GD 362. According to Figure
\ref{fg:f6}, however, our observations can rule out this last
possibility, although our limit on the helium abundance in this star
is not as stringent as for the other five objects in our
sample. Finally, we note that another target, G29-38, is a peculiar
DAZ star with a circumstellar disk. The detection of helium in this
star would have been problematic since the source of the contamination
could have been non-stellar. However, our non-detection implies that
this object can effectively be used to constrain the incomplete mixing
scenario.

We thus demonstrated for the first time that there is probably little
to no helium in the photospheres of most cool DA white dwarfs. We
believe our results put to rest the scenario invoking an incomplete
convective mixing between the hydrogen atmosphere and the underlying
helium envelope, and alternative explanations must be sought to solve
the high-$\logg$ problem in cool DA stars.

\subsection{Helium Contamination in GD 362}

For completeness and internal consistency, we reanalyze in this
section the available data for GD~362. This object represents the
coolest DA star with a direct detection of the He~\textsc{i}
$\lambda5877$ line. We reproduce in Figure \ref{fg:f7} the
observations of GD~362 from \citet{zuckerman07} taken from the Keck
Observatory Archive
(KOA)\footnote{www2.keck.hawaii.edu/inst/hires/}. We combine in our
analysis this HIRES spectrum with the lower resolution optical
spectrum of \citet{gianninas04} to determine the atmospheric
parameters ($\Te$, $\log g$, and He/H). The general fitting procedure
is described in \citet{liebert05}, although here we only use the lines
from H$\beta$ to H$\delta$ in the fit because the higher Balmer lines
are heavily contaminated by metallic lines.

Our determination of the helium abundance in GD~362 proceeds in the
same way as in \citet{zuckerman07}. The degeneracy between the He/H
abundance ratio and $\logg$ allows us to compute a sequence of best
fitting models from the Balmer line analysis, with a set of parameters
($\Te$ and $\logg$) assigned to each helium abundance. The optimal
He/H abundance ratio, and the corresponding $\Te$ and $\log g$ values,
are then found by fitting the equivalent width of the He~\textsc{i}
$\lambda5877$ line profile. Our best fitted parameters for GD 362 are
given in Figure \ref{fg:f7} together with our prediction of the helium
line profile. While the overall strength of the absorption feature is
well reproduced, the line core is not predicted as sharp and deep, a
discrepancy similar to that reported by \citet[][see their
Fig.~1]{zuckerman07}. Our value for the effective temperature,
$\Te=10,560$~K is in excellent agreement with that of
\citet[][$\Te=10,540\pm200$~K]{zuckerman07}, although both our $\log
g$ and helium abundance determinations differ, within the
uncertainties, from the values obtained by \citet{zuckerman07},
$\logg=8.24\pm 0.04$ and $\log {\rm He/H} = 1.14 \pm 0.1$ (our
internal errors are comparable to those of Zuckerman et al.). The
solution obtained by Zuckerman et al.~is actually a poor match to our
low-resolution spectrum, although the predicted helium line strength
agrees with the HIRES observations. Hence the mass we derive for GD
362, $M=0.64$ \msun\ (using evolutionary models with thin hydrogen
layers), is actually quite average (Zuckerman et al. obtained 0.73
\msun). The independent mass determination reported by
\citet{kilic08}, $M=0.74$~\msun, which is based on a measurement of
the trigonometric parallax of GD 362, actually favors the solution of
Zuckerman et al., but see our discussion below.

Since the overall procedures are similar, some of the differences
between our results and those of Zuckerman et al.~might be attributed
to our particular analysis of the higher Balmer lines, or to the
treatment of the non-ideal effects in the Hummer-Mihalas occupation
probability formalism. In any case, these atmospheric parameter
determinations must be taken with caution. First of all, the strength
of the hydrogen lines at such high helium abundances is particularly
sensitive to the Hummer-Mihalas occupation probability for neutral
particles \citep{koester05}, and a slight change in the hard sphere
radius used in the description of the neutral particle interaction may
result in significantly different atmospheric parameters. The current
parameterization (see \S~5.1) is uncertain in this high-density regime
and one could adjust the theory until there is match between our
spectroscopic solution and the constraints obtained from the
trigonometric parallax measurement of \citet{kilic08}.
\citet{gianninas04} also showed that the presence of
metals does not affect significantly the atmospheric structure of
GD~362 for pure-hydrogen atmospheres, but this assumption has not been
fully tested for helium-rich compositions. And finally, as mentioned
above, the sharp core of the helium line is not well reproduced by the
models, a feature that is not observed in the slightly hotter white
dwarf HS~0146+1847. Perhaps hydrogen is not homogeneously distributed
in the atmosphere of GD~362, especially if hydrogen is being accreted
\citep{jura09}. 

Despite the preceding discussion, it is clear that GD~362, which was
interpreted by \citet{gianninas04} as a massive ($M\sim1.24$ \msun)
white dwarf with a hydrogen-dominated and metal-rich atmospheric
composition, is in fact a more average mass white dwarf with a
helium-dominated atmosphere. According to the helium abundances
inferred in cool DA stars and displayed in Figure \ref{fg:f3}, such
objects must be rare since there are very few stars in the upper part
of this diagram. GD~362 (labeled in Fig.~\ref{fg:f3}) is actually the
object with the largest helium abundance according to our statistical
analysis. Also labeled in this figure is one of our targets, G67-23
(2246+223), which does not show any trace of helium (see
Fig.~\ref{fg:f6}). The peculiar helium-dominated star HS~0146+1847,
the hotter counterpart of GD~362, was discovered in the SPY sample
containing more than $\sim1800$ high-resolution spectra;
\citet{koester09} searched this sample for DA stars with traces of
helium, including a fairly large number with temperatures below
13,000~K. The search was also negative, but with a much larger
detection limit than provided here with our Keck I observations.
Nevertheless, it is an additional argument that such objects must be
very rare.

The existence of the peculiar atmospheric compositions of GD~362 and HS
0146+1847 may be explained by two different scenarios. We may be
witnessing white dwarfs in the process of being convectively mixed, a
process believed to occur in cooler white dwarfs
\citep{tremblay08}. This would imply, however, unusually thin hydrogen
layers according to the results shown in Figure
\ref{fg:f5}. Alternatively, the mixed H/He composition could have a
non-stellar origin, perhaps related to the collision with a water-rich
asteroid in the case of GD~362 \citep{jura09}, resulting in the
accretion of hydrogen onto a helium-rich atmosphere. The fact that
these two objects are metal-rich with circumstellar disks suggests
that the latter is the most likely explanation until we discover more
similar objects.

\section{MODEL ATMOSPHERES WITH IMPROVED STARK PROFILES}

As discussed in the introduction, our approach with respect to the
spectroscopic technique has always been to start with the simplest
white dwarf atmospheres to validate the method. For instance, BSL92
demonstrated that the spectroscopic technique is the most accurate
method for determining the atmospheric parameters of DA stars using a
sample of pure hydrogen atmosphere white dwarfs where convective
energy transport can be neglected ($\Te > 13,000$~K). B95 later
extended this spectroscopic analysis to ZZ Ceti stars, which are
located in the temperature range where the atmospheric structures are
the most sensitive to the assumed convective efficiency. They found in
particular that the ML2/$\alpha=0.6$ parameterization of the mixing
length theory (MLT) provided the best internal consistency between
optical and UV temperatures. However, this calibration of the
convective efficiency in DA stars did not change the persisting
problem that higher than average masses were found at low effective
temperatures. More recently, TB09 published Stark broadening profiles
with the first consistent implementation of the Hummer-Mihalas
equation of state. They included non-ideal perturbations from protons
and electrons directly inside the \citet{vcs} unified theory of Stark
broadening. As a first step, they revisited the BSL92 temperature
regime with improved model atmospheres of hot DA stars. They showed
convincingly that these models were adequate to describe the
observations, resulting also in $\Te$ and $\logg$ measurements
significantly larger than those obtained from previous models. In this
section, we explore the implications of these improved models on the
sepctroscopic analysis of cooler DA white dwarfs, including the ZZ
Ceti stars.

Following our approach with the spectroscopic technique, we must first
recalibrate the convective efficiency using the same approach as B95.
The tightest constraint comes from the comparison of effective
temperatures measured independently from optical and UV spectra (see
B95 for details). Other constraints, such as trigonometric parallaxes
and gravitational redshifts, are not used for this calibration; they
will be compared to our models in \S~4. We use an improved set of
$HST$ and $IUE$ near-UV observations \citep[][]{holberg2003}, some of
which were not available at the time of the B95 study. We first fit
the optical spectra to find $\Te$ and $\log g$ values. In the case of
the near-UV spectra, solutions are degenerate in $\Te$ and $\log g$
and both atmospheric parameters cannot be fitted
simultaneously. Therefore, we fit the UV spectra by forcing the
spectroscopic value of $\logg$, and the parameter $\alpha$ is then
varied until an internal consistency is achieved between UV and
optical temperatures. Figure \ref{fg:f8} (which is an updated version
of Figure 11 of B95) indicates that the ML2/$\alpha=0.8$ version of
the MLT provides the best overall internal consistency. This is a
significantly more efficient version of the MLT than estimated by B95
($\alpha=0.6$), in closer agreement with that required by nonadiabatic
models ($\alpha=1.0$) to match the empirical blue edge of the ZZ ceti
instability strip \citep[see Fig.~9 of][]{fb08}.

We have then computed extensive grids of models down to $\Te = 1500$ K
that connect with the grids used in TB09 at $\Te = 12,000$ K. Our
model grid is calculated with temperature steps of 500 K, and $\log g$
values from 6.5 to 9.5 with steps of 0.5 dex (with two additional grid
points at 7.75 and 8.25). We use the improved Stark profiles of TB09
and the updated mixing-length parameter, as discussed above. Since we
are mostly interested here in investigating the influence of our
improved profiles on the high-$\logg$ problem, we also computed a
second grid similar to that used by the Montreal group for the past
ten years. These models rely on the Stark profiles of \citet{lemke97}
with the value of the critical field $\beta_{\rm crit}$ in the
Hummer-Mihalas formalism multiplied by a factor of 2; a more detailed
discussion of this ad hoc parameter is provided in TB09. The mass
distributions for the cool DA white dwarfs in the sample of
\citet{gianninas09} are displayed in Figure \ref{fg:f9} for both sets
of model spectra. These results clearly indicate that our improved
line profiles do not help in any way to solve the long lasting problem
that the mean spectroscopic mass increases significantly below
$\Te\sim 12,500$~K. However, there is a significant improvement in the
range $30,000 > \Te > 12,500$~K, where the mass distribution appears
more constant, in contrast with our previous results where the mean
mass slightly decreases at lower temperatures, leading to an
unexpected dip in the distribution near $\Te\sim13,000$~K \citep[see
also Fig.~3 of][for a more obvious illustration of this
problem]{gianninas09}. With our improved models, we can now observe a
more stable mass distribution down to $\Te\sim12,500$~K, followed by a
sudden increase in the mean mass below this temperature.

It has been previously suggested that uncertainties in the treatment
of the Stark profiles combined with that of the pseudocontinuum
opacity originating from the dissolved atomic levels within the
occupation probability formalism of Hummer-Mihalas, and in particular
the ad hoc parameterization of $\beta_{\rm crit}$, could be
responsible for the high-$\logg$ problem \citep{koester09}. Our
results presented in Figure \ref{fg:f9} show that this is likely not
the case. The unified theory of Stark broadening from \citet{vcs}
still suffers from some approximations, but it would be surprising
that additional corrections, such as second order effects, would
change the masses by more than 1\%. Moreover, the sudden increase in
the average mass of DA stars below $\sim12,500$~K can hardly be
explained by a change in Stark profiles since these have a fairly
smooth temperature-dependence in this regime.

We finish this section by presenting additional evidence that
inaccuracies in the physics included in the model atmospheres are at
the heart of the high-$\logg$ problem. We first refer the reader to
Figure 2 of TB09, which depicts the effect of including additional
lines in the fitting procedure on the measurement of the atmospheric
parameters. The top panel shows how both $\Te$ and $\logg$ {\it
decrease} as more lines are considered in the fit. This problem has
been discussed at length by \citet{bergeron93} who proposed to
multiply by a factor of 2 the value of the critical field, $\beta_{\rm
crit}$, to partially overcome this discrepancy (see middle panel of
Fig.~2 of TB09). By including the Hummer-Mihalas formalism directly
into the line profile calculations, TB09 managed not only to get rid
of this ad hoc factor, but achieved an even better internal
consistency between the various Balmer lines (bottom panel of Fig.~2
of TB09). We show in Figure \ref{fg:f10} a similar exercise for a
typical cool DA star in our sample. The results indicate that the
$\logg$ solution shifts towards lower values, by roughly $\sim 0.5$
dex, as more lines are included in the fitting procedure. This trend
occurs for all cool DA stars in our sample, and it is observed with
all the model grids we have computed, no matter what atmospheric
composition (i.e.~mixed H/He models) or line profile theory we
assumed. We believe that the results shown here represent the ultimate
proof that best fitted models fail to properly match the spectroscopic
data.

\section{EVIDENCE FROM TRIGONOMETRIC PARALLAX MEASUREMENTS}

In this section, we look at alternative methods, independent of
spectroscopy, to measure the surface gravity --- or equivalently the
radius or the mass --- of cool DA stars. The best independent method
for measuring the atmospheric parameters of cool white dwarfs is the
photometric technique, illustrated in the left panels of Figure
\ref{fg:f2}, where optical and near-infrared photometric data are
compared with the predictions of model atmospheres \citep[see,
e.g.,][for details]{bergeron01}. With this method, only the effective
temperature and the solid angle $\pi(R/D)^2$ are considered free
parameters. When the distance $D$ is known from trigonometric parallax
measurements, one directly obtains the radius $R$ of the star, which
can then be converted into mass or $\logg$ using evolutionary
models. This method works best for cool white dwarfs since the
spectral energy distributions of hotter stars become less sensitive to
temperature. Also, since they are generally more distant, accurate
trigonometric parallaxes are more difficult to obtain.

\citet{boudreault05} used trigonometric parallax measurements for 52
DA stars to compare the photometric masses with those obtained from
the spectroscopic technique. The results (see their Fig.~1) reveal
that the spectroscopic masses are in many cases larger than the
photometric masses. \citet{bergeron07} extended this analysis by
comparing photometric and spectroscopic masses for 92 DA stars. While
the mean photometric mass of the sample appears slightly larger than
the spectroscopic mean for hot DA stars ($M\sim 0.6$ \msun; see their
Fig.~4), the dispersion of the photometric masses is significantly
larger than that obtained from spectroscopy, a result that suggests a
variable level of accuracy in the trigonometric parallax
measurements. \citet{bergeron07} also compared their spectroscopic
masses with those inferred from gravitational redshift measurements,
although the comparison remained inconclusive because of the large
uncertainties associated with the redshift velocities, which are
intrinsically more difficult to obtain than any other measurement made
with other techniques.

We conduct here a new comparison of photometric and spectroscopic
masses, but in light view of the previous studies discussed above, we
restrict our analysis to a well defined sample of homogeneous parallax
measurements for which we can better account for the uncertainties.
For instance, \citet[][and later studies]{bergeron01} relied on
trigonometric parallaxes taken from the Yale parallax catalog
\citep{ypc}. However, since this catalog represents a compilation of
different sources of parallax measurements, properly averaged with
various weights, the adopted values may exhibit a larger dispersion
than the original sources. Because it is also not clear what other
corrections are used in the Yale Catalog, we decided to go back to the
original measurements in order to build a more homogeneous sample.
The first sample discussed here is drawn from the original USNO
parallax measurements obtained from photographic plates \citep[][and
follow-up papers]{harrington80}. This sample includes 26 DA white
dwarfs with parallax uncertainties less than 12\%, and for which we
also have high signal-to-noise spectroscopy
\citep[from][]{gianninas09} as well as optical $BVRI$ and infrared
$JHK$ photometry \citep[from][]{bergeron01}. We then compute the
atmospheric parameters by using both photometric and spectroscopic
techniques. For the former method, we rely on the prescription of
\citet{holberg06} to convert the magnitudes into average fluxes. The
differences between spectroscopic and photometric masses (and
temperatures) are displayed in Figure \ref{fg:f11} as a function of
$\Te$ (photometric); known or suspected double degenerate binaries are
shown as open circles \citep[see Table 2 of][]{bergeron01}. The
temperatures obtained from both methods are generally in agreement,
within the uncertainties, with the exception of some doubles
degenerates and the two hottest stars in our sample. For the mass
comparison, no clear conclusion can be drawn for individual objects
since the observed dispersion is rather large. However, if we exclude
the double degenerates, the spectroscopic masses are larger than the
photometric masses by $\sim0.08$ \msun, on average, although the
differences appear larger for stars above 8000~K; in fact, if we
consider only stars with $\Te>8000$~K, the difference in mean mass
increases to a value of 0.11 \msun. This result can be compared to
the estimated increase in mass below $\Te=12,500$ K observed in the
top panel of Figure \ref{fg:f9}, which is of the order of 0.13 \msun,
a value entirely compatible with the mass difference observed in
Figure \ref{fg:f11}. We must therefore conclude that the photometric
masses are reasonably sound, and that the high-$\logg$ problem rests
with the spectroscopic approach.

Next, we use a more recent sample drawn from the CCD parallax survey
of nearby white dwarfs of \citet{subasavage09}. This sample contains 8
DA stars for which we perform the same analysis as above, with
$VRIJHK$ photometry (with CTIO passbands for the $R$ and $I$ filters)
and trigonometric parallaxes taken from Subasavage et al. The
comparison of the photometric and spectroscopic parameters is
displayed in Figure \ref{fg:f12}. First of all, we can see that the
high accuracy of the parallax data significantly reduces the error
bars. The object shown by the open circle is LHS 4040 (2351$-$335),
for which the $JHK$ photometry has the largest uncertainties in Table
3 of \citet{subasavage09} because of some contamination from a nearby
M dwarf (LHS 4039). If we exclude this star, the mean mass difference
is about 0.12 \msun, a value comparable to that obtained from the USNO
sample. We cannot help noticing, once again, that the largest
temperature differences occur for the hottest stars in the sample. If
we assume that the atmospheric parameters obtained from the
photometric technique are reliable, it is possible to compute model
spectra with these parameters and compare them with the observed
optical spectra. This exercise is illustrated in the right panels of
Figure \ref{fg:f13} for the three hottest stars in the Subasavage et
al.~sample, while the left panels show the best fits obtained from the
spectroscopic technique. This comparison allows us to evaluate how the
model spectra need to be corrected to yield lower spectroscopic
masses. The results suggest that the higher Balmer lines are predicted
to be too sharp by the models and the lower series are predicted to be
somewhat too narrow. Nevertheless, the differences in the model
spectra are obviously a lot more subtle than the high-mass problem
they give rise to (see Fig.~\ref{fg:f9}).

There are unfortunately very few other independent verifications of
$\logg$ or mass measurements that can be used below $\Te = 13,000$
K. More recent USNO parallaxes with CCD observations have not been
published yet, except for a very small sample \citep{monet92}. One
very interesting result with this ongoing survey is the very massive
DA star LHS 4033, for which the high mass near the Chandrasekhar limit
is predicted both from the spectroscopic and the photometric methods
\citep{dahn04}. There are also a few more recent parallax samples,
such as that of \citet{ducourant07}, but these are mostly oriented
towards very cool and faint halo white dwarfs, with very few DA stars
above 7000~K. Another way to estimate masses near the ZZ Ceti
instability strip is based on color indices that measure the strength
of the Balmer jump, which becomes sensitive to $\logg$ in this
temperature range \citep[see, e.g.,][]{WK84}. Even then, very high
precision photometry is required to measure individual
masses. \citet{koester09} applied this technique to the large SDSS
sample to compute the average photometric $\logg$ value as a function
of temperature (see their Fig.~2). On average, the photometric $\logg$
values are not shown to rise significantly in the range of 16,000 K $<
\Te <$ 9000 K, pointing again to a problem with the spectroscopic
approach.

\section{ALTERNATIVE SOLUTIONS TO THE HIGH-LOG G PROBLEM}

In the remainder of this paper, we explore alternative solutions to
the high-$\logg$ problem in cool DA stars. Some of these solutions
have already been discussed by \citet{koester09}, but we present here
some additional arguments. The first thing to realize is that with the
exception of the sudden increase in $\logg$ below $\Te\sim 12,500$~K,
the {\it dispersion} of $\logg$ values in a given range of
temperatures remains nearly identical down to $\Te = 8000$~K (see
Fig.~\ref{fg:f1}). Therefore, the solution we are looking for is
expected to impact {\it all} cool DA white dwarfs in a systematic
fashion. We begin by showing in panel (a) of Figure \ref{fg:f14} the
shifts in $\logg$ as a function of effective temperature required to
obtain a mean mass comparable to that of hotter DA stars. These shifts
have been estimated by computing, in 1000~K temperature bins, the mean
$\logg$ value for the sample displayed in Figure \ref{fg:f1}. Then,
to estimate the effects of each experiment (described below) on the
spectroscopic determinations of $\Te$ and $\logg$, we simply fit our
reference grid of model spectra with our test grid, and report the
results in the various panels of Figure \ref{fg:f14}.

\subsection{Hummer \& Mihalas equation-of-state: Neutral Interactions}

The modeling of cool and dense plasma for a non-ideal gas --- mostly
volume effects accounting for the finite size of the neutral particles
that can lead to pressure ionization at higher densities --- has
always remained one of the most difficult areas in
physics. Fortunately, non-ideal effects due to neutral particles are
fairly unimportant for the thermodynamical structure of hydrogen-rich
white dwarf atmospheres, and they only have an impact on the excited
states of hydrogen. For these applications, the simple Hummer-Mihalas
occupation probability model, in which the atoms are considered as
hard spheres with a constant radius, has been used in most white dwarf
atmosphere codes. In this theory, the atomic states are effectively
destroyed when the distance between two particles is smaller than the
corresponding atomic radii. The radius of the hydrogen hard sphere is
usually taken as $n^2a_0$ where $n$ is the principal quantum number
and $a_0$ is the Bohr radius. This is obviously a crude approximation
that does not take into account interaction
potentials. \citet{bergeron91} found that a direct implementation of
the occupation probability formalism yields $\logg$ values that are
too low in the regime where non-ideal effects become dominated by
neutral interactions ($\Te\lesssim 8000$~K). The problem is that the
higher Balmer lines are predicted too weak at normal surface
gravities. Therefore, Bergeron et al.~divided the hydrogen radius by a
factor of two to reduce the non-ideal effects for the higher lines of
the series. In view of the preceding discussion of the simple model
used to describe neutral interactions, a 50\% change of the effective
atomic radius is not physically unrealistic.

One problem with the parameterization of the hard sphere model used by
\citet{bergeron91} is that, since $\logg$ values are already too high
for all DA stars below $\Te \sim 12,500$ K, it is not clear what value
of the mean surface gravity should be used to ``calibrate'' the hard
sphere radius. In view of these uncertainties, we simply show in panel
(b) of Figure \ref{fg:f14} how the atmospheric parameters are changed
if we use the original parameterization of \citet{hm88} where the
hydrogen radius is set to its standard value ($n^2a_0$). The effects
are small at temperatures higher than $\Te\sim 10,000$ K. However, if
we increase the hydrogen radius even more, the atmospheric parameters
become totally unrealistic at low temperatures. The conclusion, which
was understood early-on (see \S~2.6 of B95), is that an inaccurate
treatment of the hard sphere model in the occupation probability
formalism cannot be the source of the high-$\logg$ problem.

We finally note that there is another discrepancy in the
Hummer-Mihalas formalism for neutral interactions, which predicts ---
in cool hydrogen-atmosphere white dwarfs --- a largely overestimated
contribution of the bound-free opacity from the Lyman edge associated
with the so-called dissolved atomic levels of the hydrogen atom, or
pseudocontinuum opacity \citep{kowalski06a}. This is partially because
the theory fails to take into account the interaction potentials for
the H-H and H-H$_2$ collisions. More realistic calculations
\citep{kowalski06b} actually show that this opacity source is
completely negligible in cool DA stars and should have no effect on
the spectroscopic results discussed here.

\subsection{Neutral Broadening}

The broadening profiles of hydrogen lines in cool DA stars become
increasingly dominated by neutral broadening, for which resonance
broadening is the dominant source. It has traditionally been taken
into account by using broadening parameters from \citet{ali65}, but
new theories \citep{barklem00} claim that these parameters could be in
error by a factor of two, although the exact numbers are still being
disputed \citep{allard08}. In panel (c) of Figure \ref{fg:f14}, we
show how the atmospheric parameters are affected when we multiply
arbitrarily the resonant broadening parameter by a factor of two. Not
surprisingly, the temperature range where the effect is observed is
similar to that of the non-ideal gas effects due to neutral particles,
shown in panel (b). Therefore, even if neutral broadening was much in
error, any improvement could not account for a shift in the mean
$\logg$ values at higher temperatures.

\subsection{Hummer \& Mihalas equation-of-state: Charged Particle Interactions}

TB09 presented the first model spectra for DA stars that took into
account in a consistent way the non-ideal effects described in the
Hummer-Mihalas theory both in the occupation probability and line
profiles calculations. The non-ideal effects due to protons are fairly
well understood and we can safely say that, with the help of plasma
experiments, these effects are well described to within a few percent,
especially for the hydrogen Balmer lines. One matter discussed in TB09
is that the Hummer-Mihalas formalism originally used the Holtsmark
distribution to describe the proton microfields. It was later upgraded
with the \citet{hooper68} distribution \citep{nayfonov99}, including
the Debye shielding effect. In panel (d) of Figure \ref{fg:f14}, we
show that even by using the more general and precise microfield
distribution of \citet{potekhin02}, differences in the models remain
extremely small.

It must be understood, however, that the Hummer-Mihalas occupation
probability formalism is based on a dipole interaction between the
absorber and the protons, an approach that fails for very close
collisions. Such collisions are better described with a proper account
of the colliding particle potential (see TB09 for details). Proper
calculations show that the non-ideal effects described in \citet{hm88}
cannot lower the ionization potential to energies that correspond
usually to transitions between the first two levels. Therefore, the
bound-free pseudocontinuum opacity from the Lyman edge can be
considered negligible near the L$\alpha$ region. The transition to the
Hummer-Mihalas theory at lower wavelengths is still not well
understood, and a sharp cutoff of the Lyman pseudocontinuum opacity
longward of L$\gamma$ ($\lambda > 972.5$ \AA) is currently used in our
models. However, as seen in panel (e) of Figure \ref{fg:f14}, even if
we extend our arbitrary cutoff to L$\alpha$ ($\lambda > 1216$~\AA),
the impact on the Balmer line analysis remains completely negligible.

Finally, the Hummer-Mihalas theory also takes into account non-ideal
effects due to electronic collisions. These have never been included
in white dwarf model atmospheres until the work of TB09. In panel (f)
of Figure \ref{fg:f14}, we show how the atmospheric parameters are
changed if this additional contribution to the occupation probability
is removed from {\it both} the line profile and equation-of-state
calculations. While these non-ideal effects cannot be neglected
altogether, they have little influence on the atmospheric parameters
determined from spectroscopy. Our conclusion for this section is that
the high-$\logg$ problem cannot be easily explained by some inaccurate
treatment of non-ideal effects due to charged particles.

\subsection{H$^{-}$ Continuum Opacities}

The dominant source of opacity in cool DA atmospheres is the
bound-free and free-free H$^{-}$ opacity. The main uncertainty in the
calculation of the H$^{-}$ opacity is the approximation used in the
wave function expansions, although the current accuracy is believed to
be better than 1\% \citep{john88}. In panel (g) of Figure
\ref{fg:f14}, we artificially increased the contribution of the
H$^{-}$ opacity in our models by 15\%. The results indicate that such
uncertainties would have absolutely no effect on the spectroscopic
determination of the atmospheric parameters of cool DA stars.

\subsection{Convective Energy Transport}

We discussed in our introduction that the use of the mixing-length
theory to treat the convective energy transport in white dwarf
atmospheres may represent a significant source of physical uncertainty
in the calculations of the atmospheric structures. It has even been
considered the only viable solution to the high-$\logg$ problem by
\citet{koester09}. Using our improved model atmospheres, we proceeded
in \S~3 to a recalibration of the MLT free-parameter $\alpha$ that
describes the convective efficiency. None of the discrepancies
presented in this work provide any {\it direct} evidence that
something is wrong with the MLT framework. The only circumstantial
evidence is that the sudden increase in $\logg$ values below
$\Te\sim12,500$~K coincide precisely with the temperature where the
extent of the hydrogen convection zone becomes significant (see
Fig.~\ref{fg:f5}). It is also the temperature range where our models
predict that a majority of the energy flux is transported by
convection in the photosphere ($\tau_R=1$). Since convective energy
transport may change dramatically the temperature and pressure
structures of the atmosphere, the predicted line profiles can be
affected significantly since different parts of the line profiles are
formed at different optical depths.

Our first objective is to evaluate the sensitivity of the high-$\logg$
problem to the current MLT models, and investigate whether there is
any way to solve the problem within the theory. We previously
discussed in \S~3 that the spectroscopic technique yields different
solutions for $\Te$ and $\logg$ depending on the particular choice of
the free parameter $\alpha$. This is why we constrained the
convective efficiency by relying on independent near-UV observations.
If we ignore this constraint, it is possible to achieve reasonable
fits in the optical using any value of $\alpha$. However, B95 showed
(see their Fig.~3) that the distribution of DA stars as a function of
effective temperature shows unexpected gaps or clumps both in the
limit of low and high convective efficiencies. This is because the
hydrogen Balmer lines, which reach their maximum strength near $\sim
13,000$~K, are predicted too strong or too weak, respectively. In
panel (h) of Figure \ref{fg:f14}, we show the effect of adopting a
value of ML2/$\alpha=1.5$ (instead of 0.8) in our model
calculations. The surface gravities are indeed lowered in the range
$13,000 >\Te> 10,000$~K, but as discussed above, the distribution of
DA stars now shows an important clump in temperature since the Balmer
lines are predicted too weak near $\Te\sim 13,000$~K (see Fig.~4 of
B95), and DA stars with the strongest lines accumulate in this region.

Even with a significantly increased convective efficiency, the $\logg$
values below 10,000~K remain too high. A variation of the convective
efficiency in this particular range of temperature has little effect
on the predicted line profiles, and the spectroscopic determinations
of $\logg$ become somewhat independent of the particular choice of the
parameter $\alpha$. Indeed, convection becomes increasingly adiabatic
in this range of effective temperature, and thus the convective flux
becomes independent of the MLT. The temperature gradient at
$\tau_R=1$, however, is still significantly different from the
adiabatic gradient ($>5\%$) in models with $\Te
\sim 8000$ K. Therefore, there is still room for another convection
theory that could yield different temperature profiles in the range
$10,000 >\Te> 8000$~K. In other words, the fact that the $\logg$
values become independent of the particular choice of $\alpha$ does
not guarantee that we converged to the correct solution. The only
conclusion we can reach so far is that {\it within} the MLT framework,
we are unable to solve the high-$\logg$ problem.

As reviewed by \citet[][see their \S~4.1]{fb08}, nonadiabatic
calculations require a higher convective efficiency ($\alpha=1.0$) to
account for the empirical blue edge of the ZZ Ceti instability
strip. Taken at face value, this conclusion implies that convection in
the atmospheric layers is less efficient than in the deeper layers.
From 2D hydrodynamic simulations, \citet{ludwig94} also concluded that
the convective efficiency must increase with depth, with a value of
$\alpha$ at $\tau_R\sim10$ in the range of 4 to 5. In light of these
findings, we performed a final test by using a variable mixing-length
as a function of depth, keeping the current value at
$\tau_R=1$.\footnote{More specifically, we used a linear variation
with depth from $\alpha=0.8$ at $\tau_R=1$ to $\alpha=3.2$ at
$\tau_R =100$.} The results, not shown here, reveal that the effects
produced on the temperature structure deep in the atmosphere have
absolutely no impact on the predicted spectra. Therefore, while an
increase in convective efficiency with depth may affect the
predictions from pulsational models, it has little effect on model
spectra of cool DA stars.

It seems we may now have reached the limit of the 1D mixing-length
theory, and one is naturally looking at solutions from more realistic
radiation hydrodynamic calculations, such as the 2D simulations of
\citet{ludwig94}. These calculations have not been followed up,
unfortunately, since 3D stellar atmospheres with radiation
hydrodynamics have become more stable and accurate
\citep{nordlund09}. From a limited set of models representing a
typical ZZ Ceti star, \citet{ludwig94} concluded that the
mixing-length theory is roughly correct. However, since the
differences between the current model spectra and the expected spectra
are rather small (see the right panels of Fig.~\ref{fg:f13}), this
conclusion about the validity of the MLT framework should be taken
with caution since the calculations of Ludwig et al.~have not been
confronted to real stellar spectra. A complete grid of 2D, or even 3D
models should be compared with observations before reaching this
conclusion. This represents the obvious next step in attempting to
solve the high-$\logg$ problem.

\section{CONCLUSION}

The starting point of our study was the longstanding problem that
spectroscopic values of $\logg$ --- or masses --- of DA white dwarfs
cooler than $\Te\lesssim12,500$~K are significantly higher than those
of hotter DA stars, a feature observed in {\it all spectroscopic
analyses} published to date. This discrepancy limits our ability to
accurately measure the atmospheric parameters of the vast majority of
DA white dwarfs older than $\sim 1$ Gyr, and may affect the results of
ZZ Ceti pulsation analyses, as well as our ability to use white dwarf
stars as precise cosmochronometers or distance indicators. The first
solution proposed to explain the high-$\logg$ problem has been to
recognize that a mild and systematic helium contamination from
convective mixing could mimic the effects of high surface gravities.
In this paper, we presented high signal-to-noise, high-resolution
spectroscopic observations obtained with the Keck I telescope of the
region near the He~\textsc{i} $\lambda5877$ line for six cool DA white
dwarfs. {\it We did not detect helium in any of our target stars.}
Upper limits on the helium abundance in five of our targets allowed us
to rule out the incomplete convective mixing scenario as the source of
the high-$\logg$ problem. This result is in line with the conclusions
of \citet{tremblay08} that most DA white dwarfs ($\sim$85\%) have
thick hydrogen layers ($M_{\rm H}/M_{\rm tot} >10^{-6}$). In view of
our results, we revisited the nature of the high-$\logg$ problem and
presented evidence based on photometric masses that the conundrum lies
with the spectroscopic technique itself. We showed in this context
that the improved Stark profiles of \citet{tremblay09} for the
modeling of cool DA spectra did not help to solve the problem.

In our review of alternative solutions, we concluded --- as
\citet{koester09} did --- that a problem with the treatment of
convective energy transport, currently the mixing-length theory, is
the most plausible explanation for the high-$\logg$ problem. While
nothing points to a specific discrepancy with the MLT framework, the
next obvious step is to compare our current atmospheric parameters for
DA stars with those eventually obtained from more detailed
hydrodynamic models. Only then will we be able to confirm this
hypothesis.

\acknowledgements
This work was supported in part by the NSERC Canada and by the Fund
FQRNT (Qu\'ebec). P.B. is a Cottrell Scholar of Research Corporation for
Science Advancement. Some of the data presented here were obtained at
the W.~M. Keck Observatory, which is operated as a scientific
partnership among the California Institute of Technology, the
University of California and the National Aeronautics and Space
Administration. The Observatory was made possible by the generous
financial support of the W.~M. Keck Foundation. This research has made
use of the Keck Observatory Archive (KOA), which is operated by the
W.~M. Keck Observatory and the NASA Exoplanet Science Institute
(NExScI), under contract with the National Aeronautics and Space
Administration.

\clearpage
\clearpage
\begin{deluxetable}{clcccc}
\tablecolumns{6}
\tablewidth{0pt}
\tablecaption{Upper Limits on the Helium Abundance in Cool DA White Dwarfs}
\tablehead{
\colhead{WD} &
\colhead{Name} &
\colhead{$\Te$~(K)$^{\rm a}$} &
\colhead{$\logg^{\rm a}$} &
\colhead{He/H (3$\sigma$)} &
\colhead{Note}
}
\startdata
 0133$-$116 &Ross 548  & 12470  & 8.05 & $<$0.04 & 1\\        
 0415$+$271 &HL Tau 76 & 11780  & 7.99   &$<$0.11 & 1\\    
 1855$+$338 &G207-9  & 12390  & 8.42   &$<$0.09 & 1\\  
 1935$+$276 &G185-32 & 12630  & 8.13   &$<$0.04 & 1\\  
 2246$+$223 &G67-23 & 10720  & 8.89&$<$1.42 & \\  
 2326$+$049 &G29-38 & 12200  & 8.22   &$<$0.04& 1,2\\  
\enddata
\tablenotetext{a}{$\Te$ and $\log g$ assume pure hydrogen models.}
\tablecomments{(1) ZZ Ceti stars; (2) DAZ star with a circumstellar disk.}
\end{deluxetable}

\clearpage

\figcaption[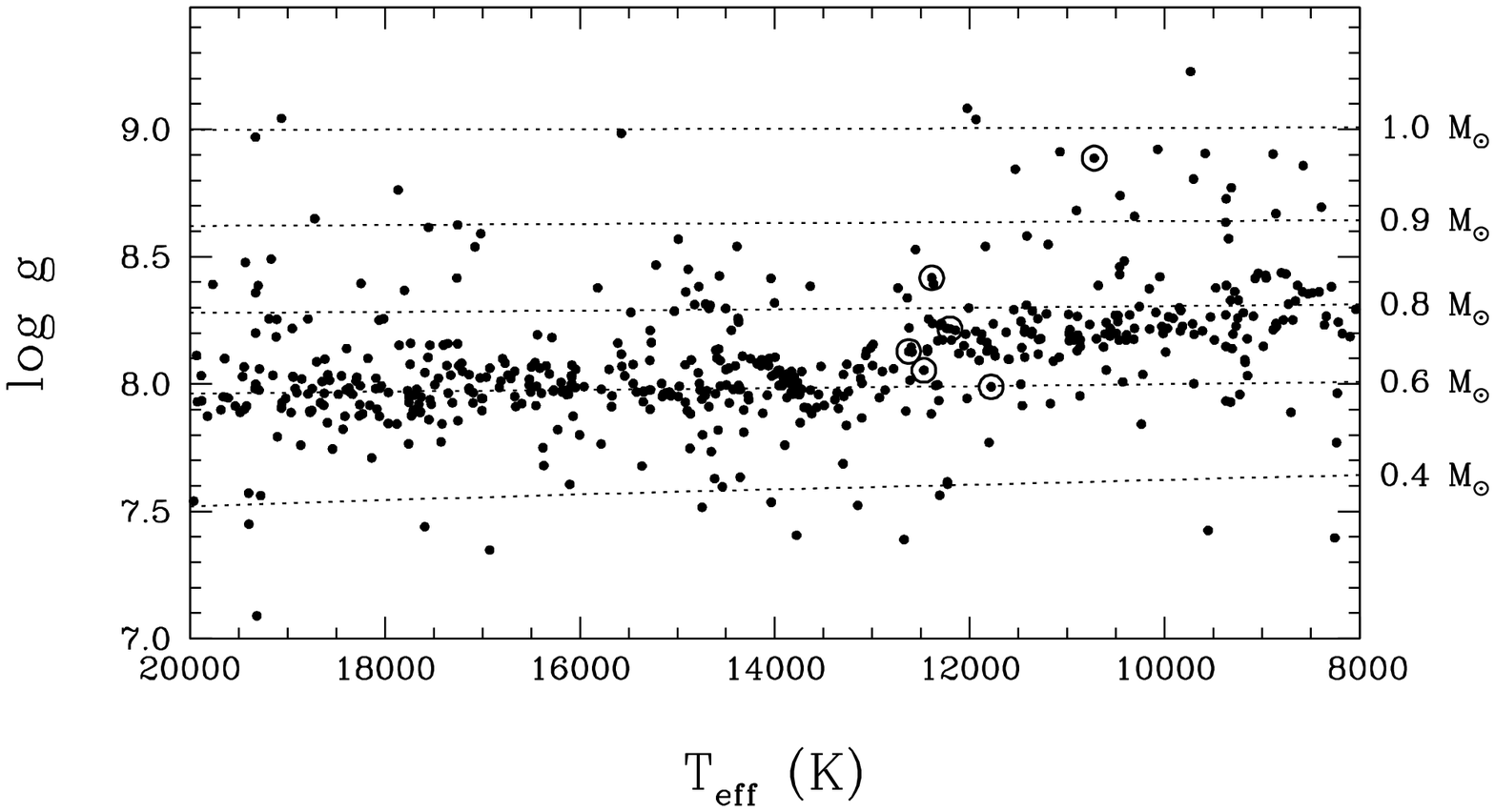] 
{Distribution of $\logg$ as a function of effective temperature for DA
white dwarfs drawn from the sample of \citet{gianninas09}. The
atmospheric parameters are determined using the reference model grid
described in this work (see \S~3), which is based on the improved
Stark broadening profiles of TB09. Evolutionary models at constant
mass from \citet{fontaine01} are shown with dashed lines and
identified on the righthand side. The circles indicate the location of
the six white dwarfs observed with HIRES on the Keck I 10-m telescope,
and identified in Table 1.
\label{fg:f1}}

\figcaption[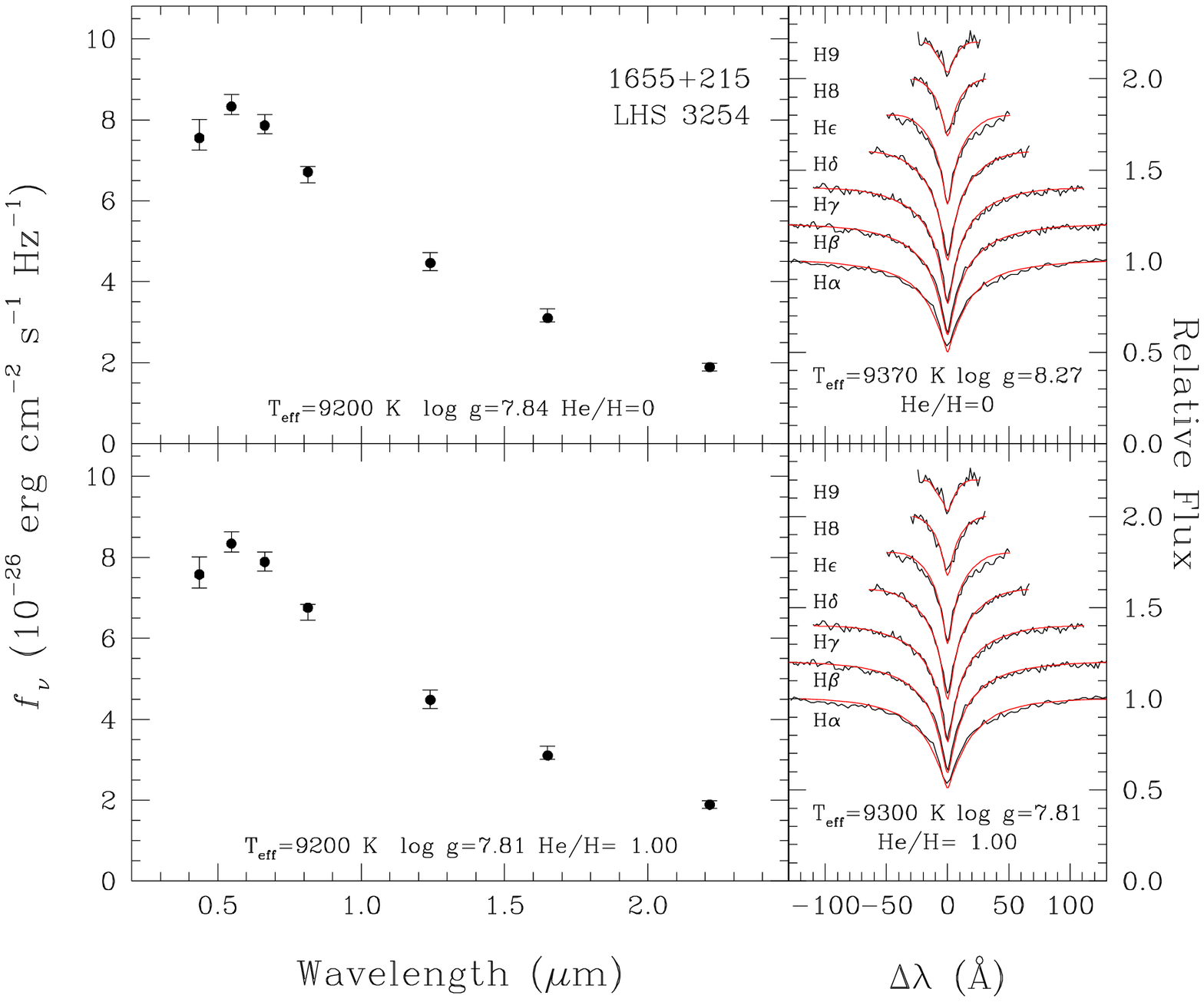] 
{Our best spectroscopic and photometric fits to the DA star LHS 3254
(1655+215) using pure hydrogen models ({\it top panels}). In the left
panel, the error bars represent the observed fluxes derived from
optical $BVRI$ and infrared $JHK$ magnitudes, while the corresponding
model fluxes are shown as filled circles. The 0.43 dex discrepancy in $\logg$
between both fitting techniques can be resolved if helium is
allowed in the atmosphere ({\it bottom panels}). From a qualitative
point of view, both the pure hydrogen and mixed He/H solutions are
indistinguishable. This is an updated version of Figure 20 from
\citet{bergeron01}.
\label{fg:f2}}

\figcaption[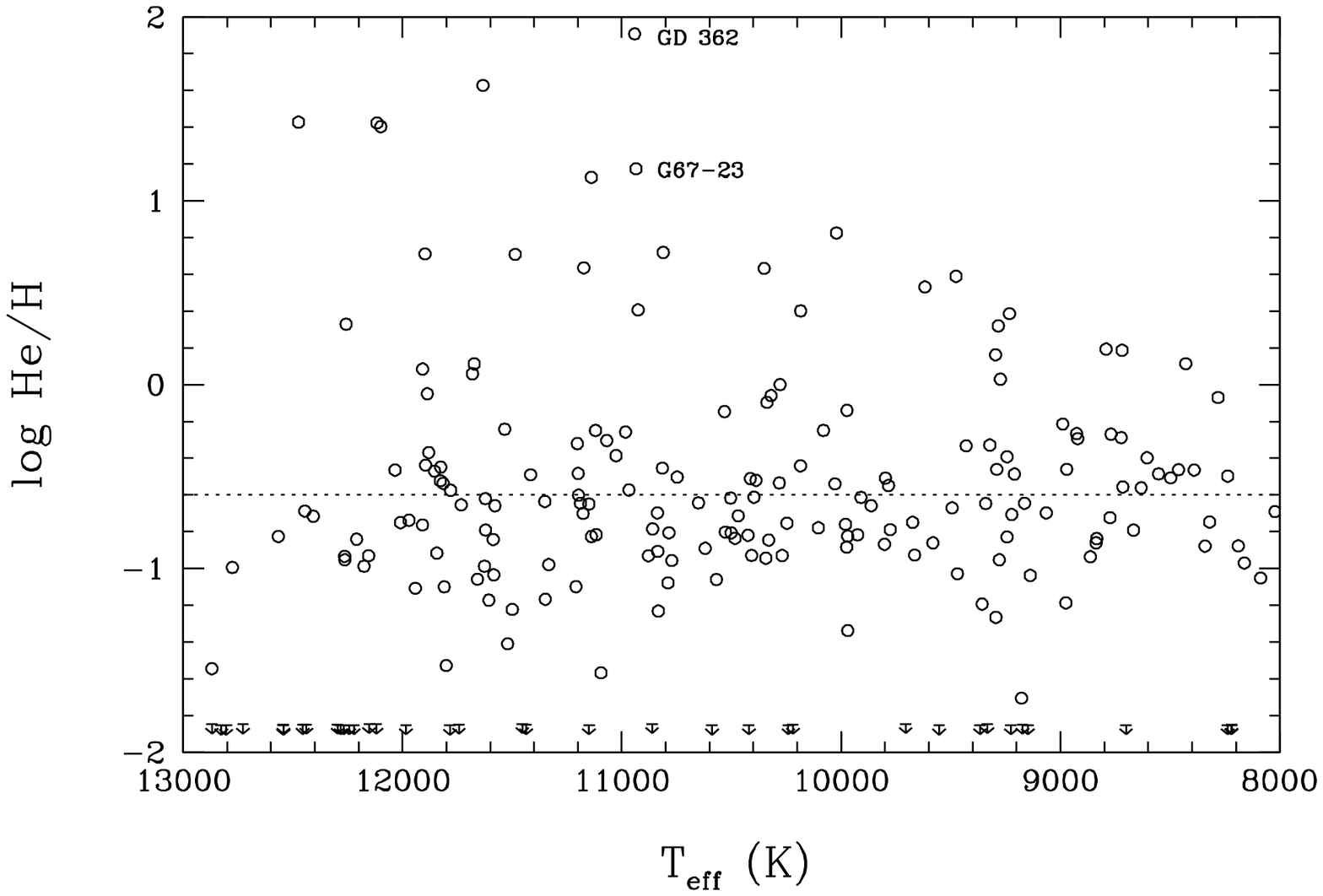] 
{Helium abundance determinations as a function of $\Te$ for the cool
DA white dwarfs in the sample of \citet[][same as
Fig.~\ref{fg:f1}]{gianninas09}, assuming a mass of 0.649 \msun\ for
all stars. The two objects labeled in the figure are discussed in the
text. The downward arrows at the bottom of the figure represent the
objects for which the spectroscopic mass is already smaller than 0.649
\msun. The dashed line indicates a systematic contamination of He/H =
0.25 used in our discussion.
\label{fg:f3}}

\figcaption[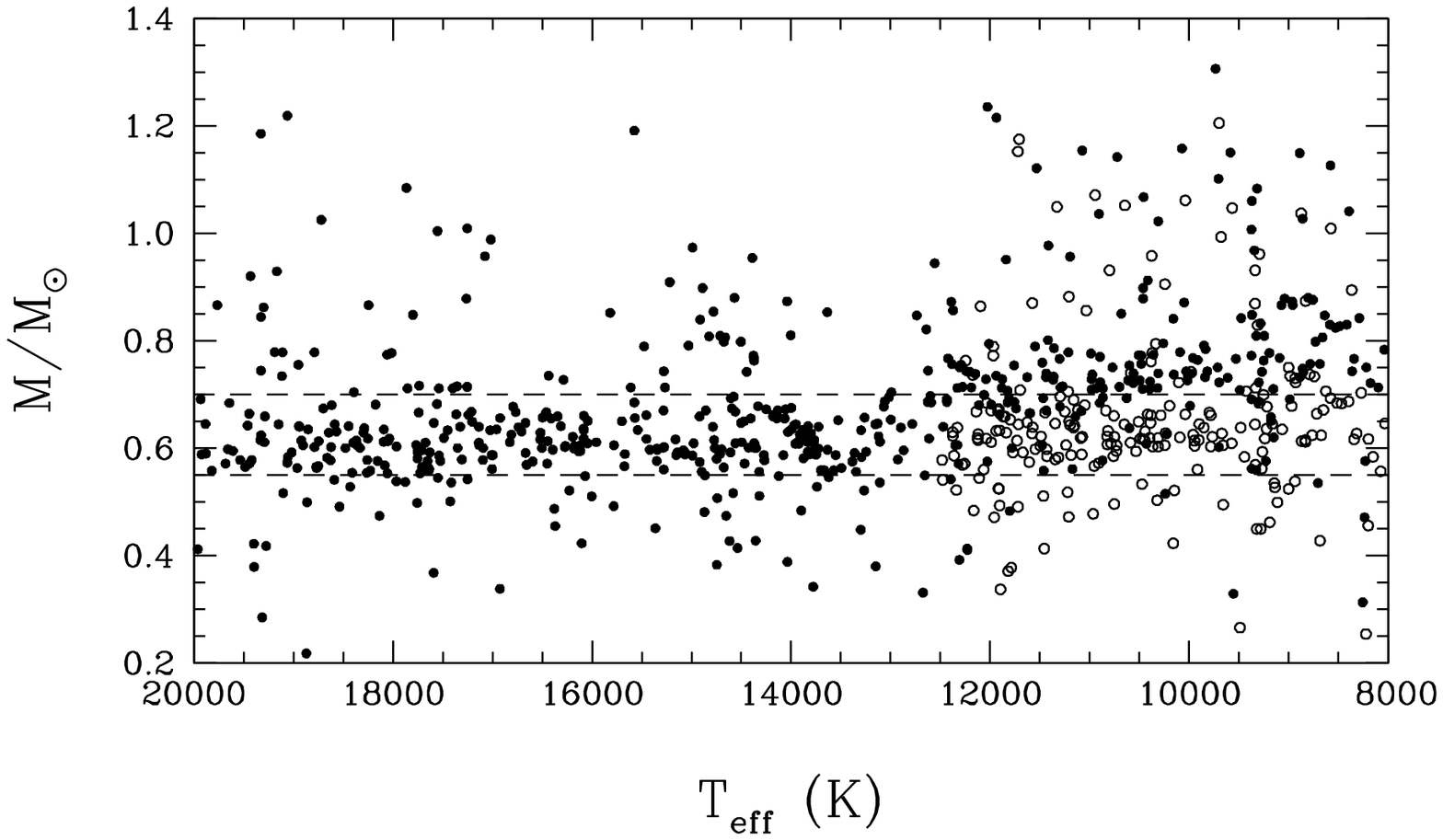] 
{Filled circles: same as Figure \ref{fg:f1} but with the $\logg$
values converted into mass using the evolutionary models of
\citet{fontaine01} with thick hydrogen layers; lines of constant 
mass at 0.55 and 0.70 \msun\ are shown as a reference. Open circles:
mass distribution of white dwarfs below $\Te=12,500$~K obtained from
model spectra that contain small traces of helium with He/H =
0.25. Such a mild but systematic helium contamination produces a
consistent mass distribution throughout the entire temperature range
displayed here.
\label{fg:f4}}

\figcaption[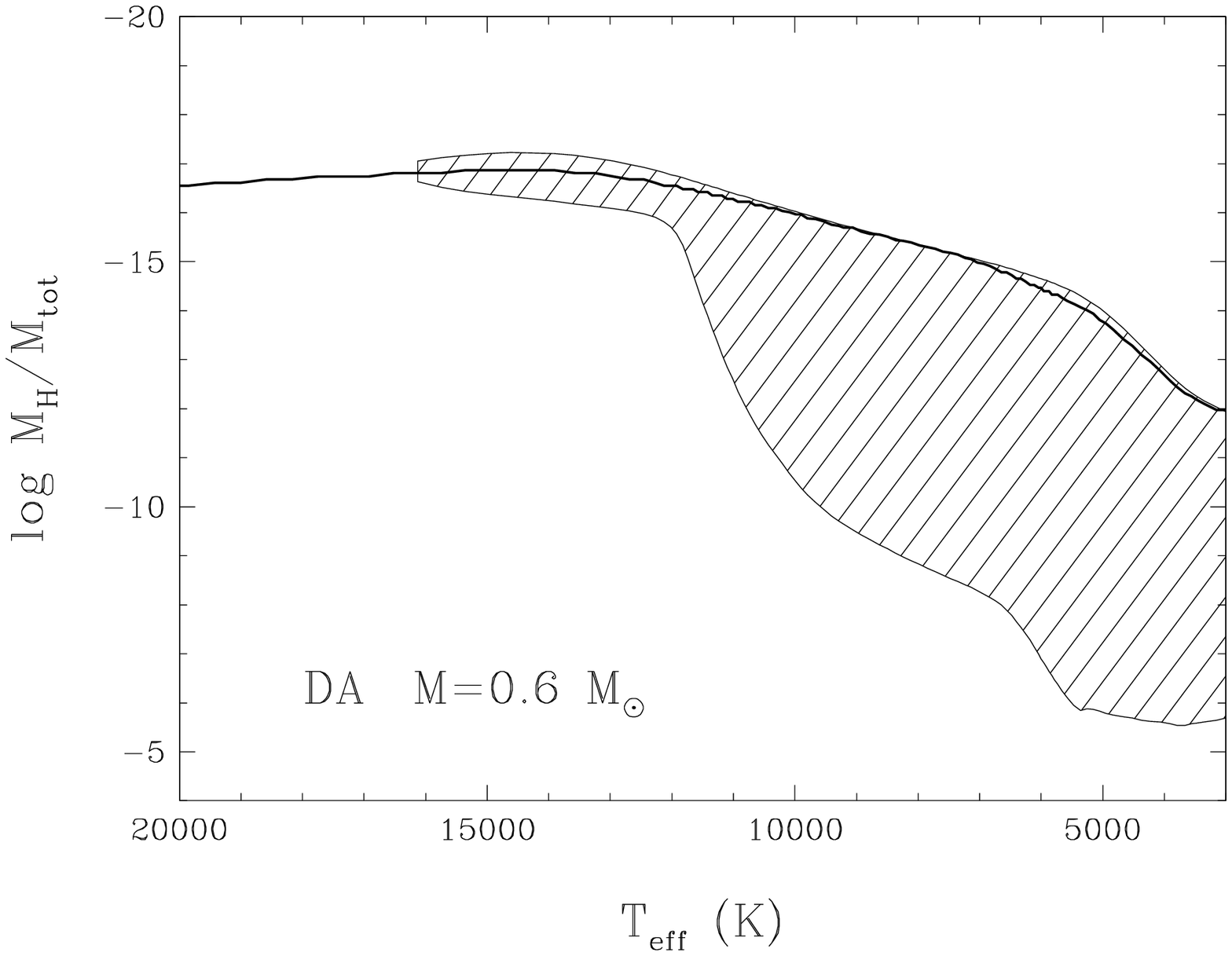] 
{Location of the hydrogen convection zone ({\it hatched region}) as a
function of effective temperature in the pure hydrogen envelope of a 0.6
\msun\ DA white dwarf calculated with the ML2/$\alpha=0.6$ version of the
mixing-length theory (from G.~Fontaine \& P.~Brassard 2006, private
communication). The depth is expressed as the fractional mass above
the point of interest with respect to the total mass of the star. The
thick solid line corresponds to the photosphere ($\tau_R\sim1$).
\label{fg:f5}}

\figcaption[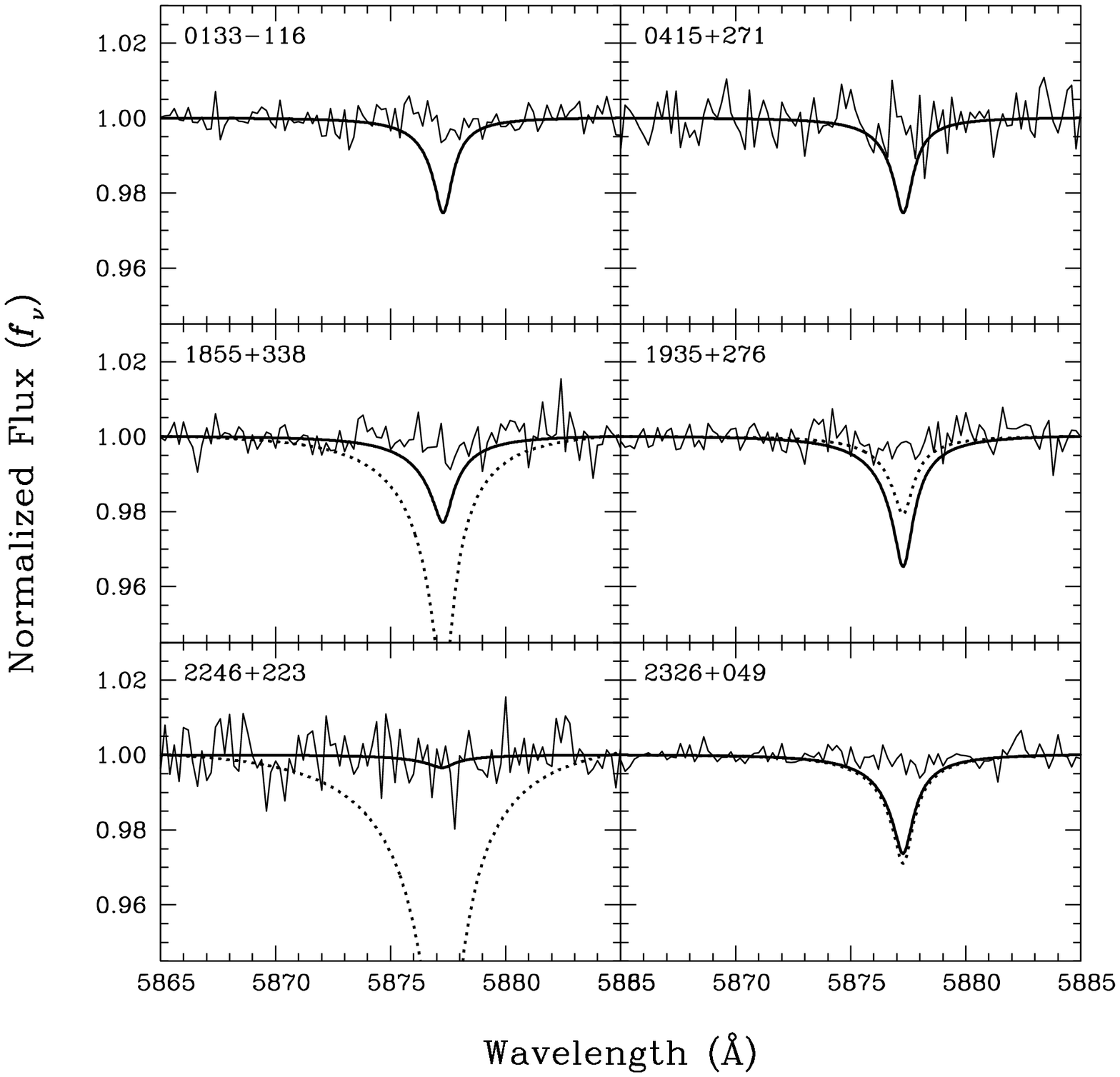] 
{High-resolution spectra for the six white dwarfs observed with the
HIRES spectrograph on the Keck I 10-m telescope. Predictions from
model atmospheres with a systematic helium contamination of He/H =
0.25 are shown as thick solid lines. For four objects (1855+338,
1935+276, 2246+223, and 2326+049), we also show as dotted lines the
predicted spectra for an assumed mass of 0.649 \msun\ and helium
abundances determined from our spectroscopic fits to low-resolution
spectra (the two other objects have spectroscopic masses already lower
than 0.649 \msun). Both the observed and model spectra are shown with a 0.2
\AA~resolution for clarity.\label{fg:f6}}

\figcaption[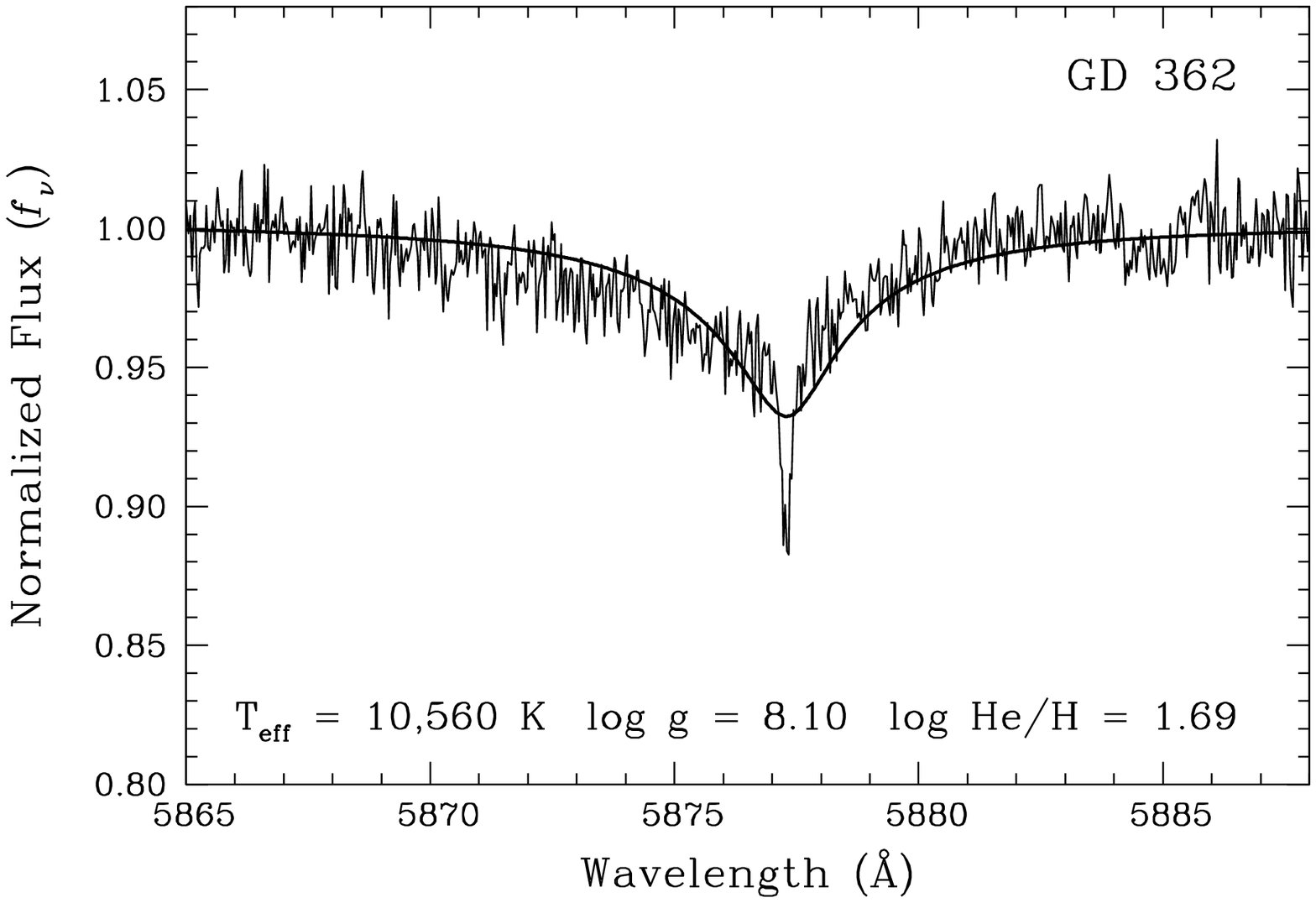] 
{The observed HIRES spectrum of GD~362 \citep{zuckerman07} from the
Keck Observatory Archive. The raw data were reduced using MAKEE
and calibrated to vacuum wavelength. We added the six
similar exposures obtained over three different nights. The
spectrum is used to constrain the He/H ratio using the best fit to the
equivalent width of the line. The $\Te$ and $\log g$ values are
determined from a fit to our low-resolution Balmer spectrum with mixed
He/H models. The resulting atmospheric parameters are given in the
figure.
\label{fg:f7}}

\figcaption[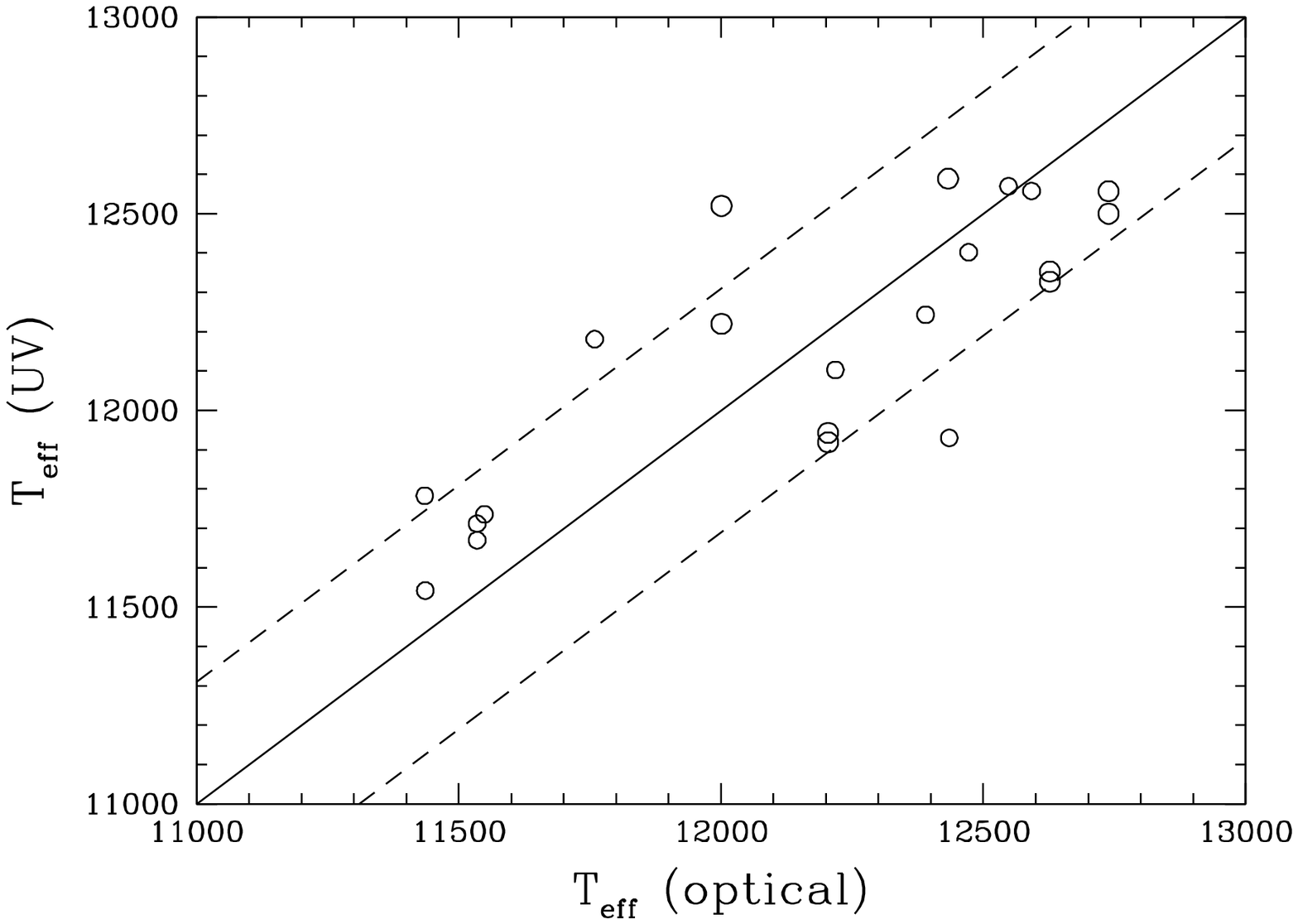] 
{Effective temperatures of ZZ Ceti stars derived from UV spectra
compared with the optical determinations. Optical $\logg$ values are
assumed in the determination of the UV temperatures. The size of the
symbols reflects the different weights assigned to the UV spectra, as
discussed in B95. The solid line indicates the locus where $\Te({\rm
optical})=\Te({\rm UV})$, while the dashed lines represent the $\pm
350$~K uncertainty allowed by the optical analysis. The results
indicate that model atmospheres calculated with the ML2/$\alpha=0.8$
parameterization of the mixing-length theory provide the best internal
consistency between optical and UV temperatures.
\label{fg:f8}}

\figcaption[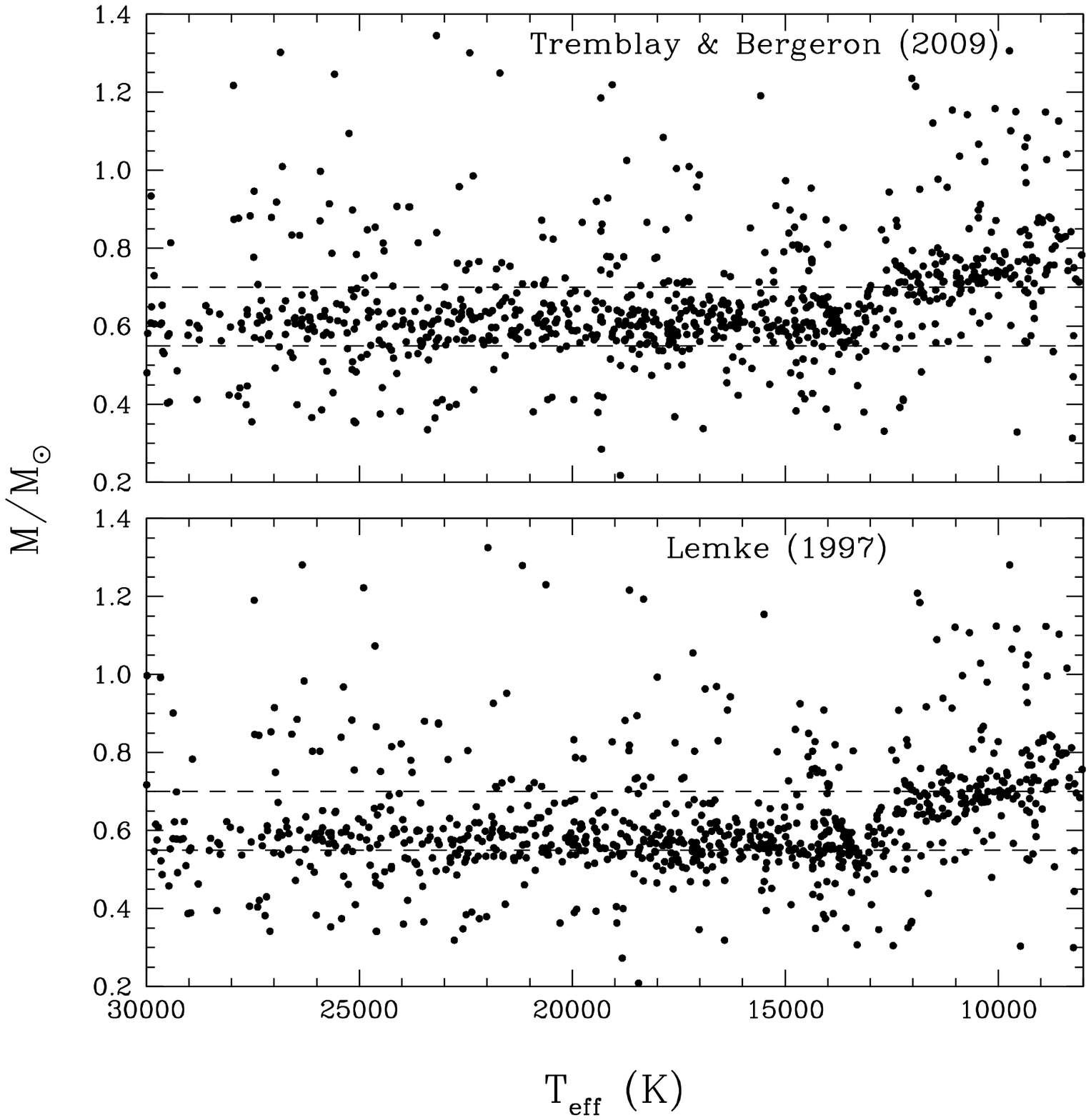] 
{Mass distribution as a function of $\Te$ for the DA white dwarfs in
the sample of \citet{gianninas09}. The atmospheric parameters are
derived from spectroscopic fits using model spectra calculated with
the Stark broadening profiles of TB09 ({\it top panel}) and with the
Stark profiles from \citet{lemke97} and the value of the critical
field ($\beta_{\rm crit}$) multiplied by factor of 2 ({\it bottom
panel}). Lines of constant mass at 0.55 and 0.70 \msun\ are shown as
a reference.
\label{fg:f9}}

\figcaption[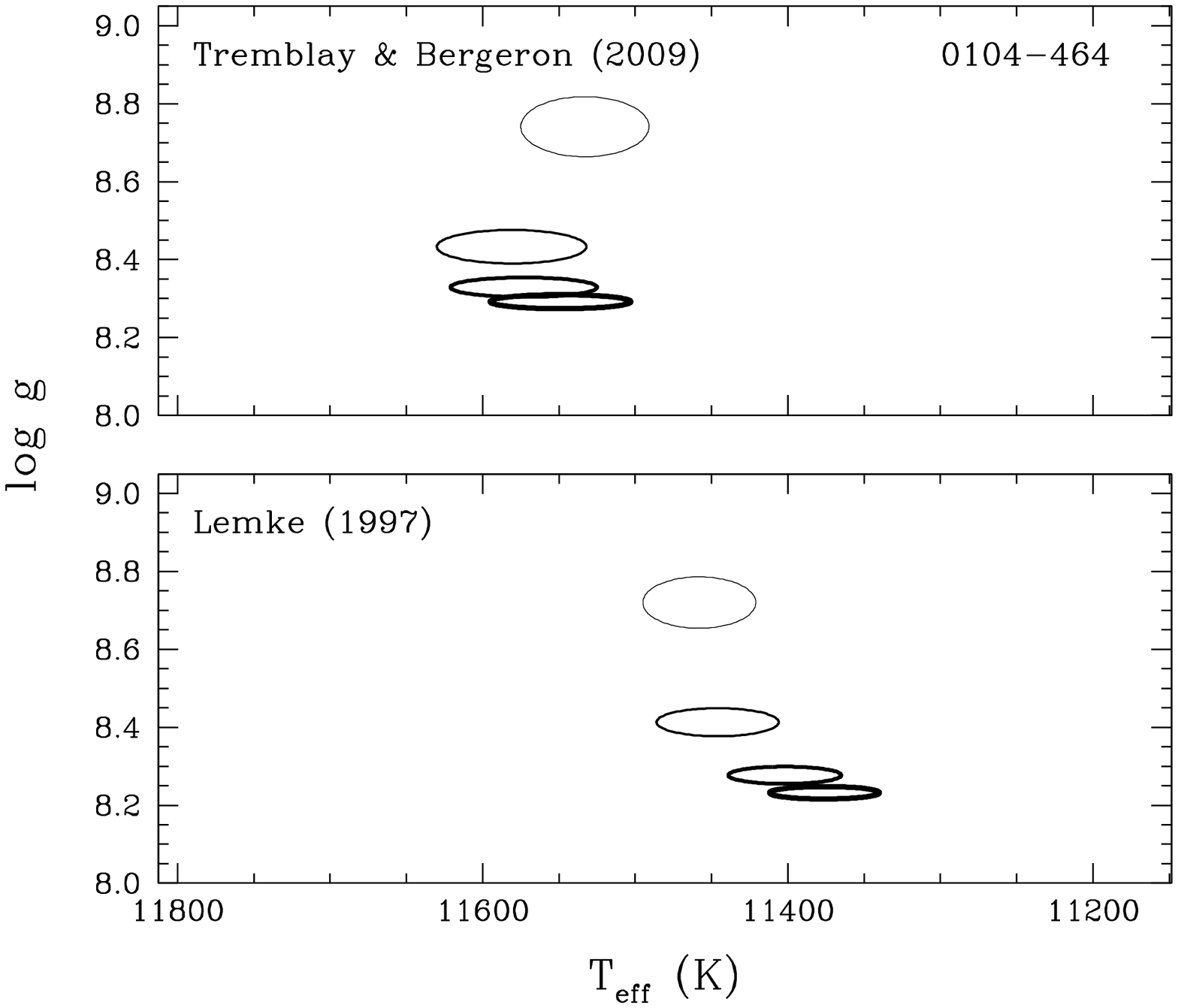] 
{Solutions in a $\Te - \logg$ diagram for a ZZ Ceti star using 2, 3, 4
and 5 lines (H$\beta$ up to H8) in the fitting procedure (represented
by progressively thicker 1 $\sigma$ uncertainty ellipses from our fitting
procedure). The different panels show the results for the two grids
discussed in the text. 
\label{fg:f10}}

\figcaption[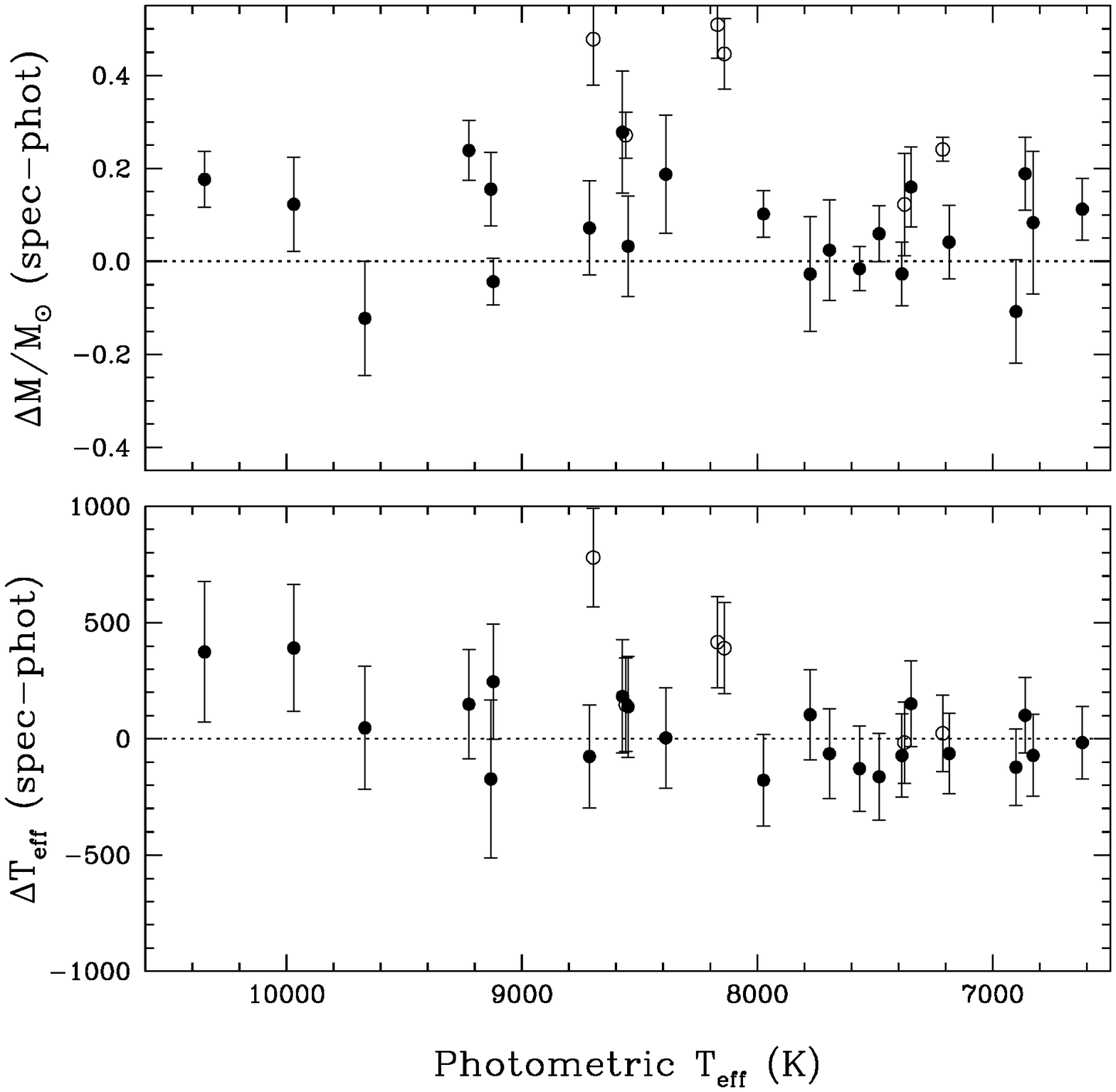] 
{Differences between white dwarf parameters (effective temperature and
mass) determined from the spectroscopic and photometric methods as a
function of photometric temperatures. The $BVRIJHK$ photometric data
used in these determinations are taken from \citet{bergeron01} with
trigonometric parallaxes from the USNO catalogs, while the optical
spectra are from our own archive. The open circles represent suspected
or known double degenerates. The horizontal dotted lines correspond to
a perfect match between spectroscopic and photometric parameters.
\label{fg:f11}}

\figcaption[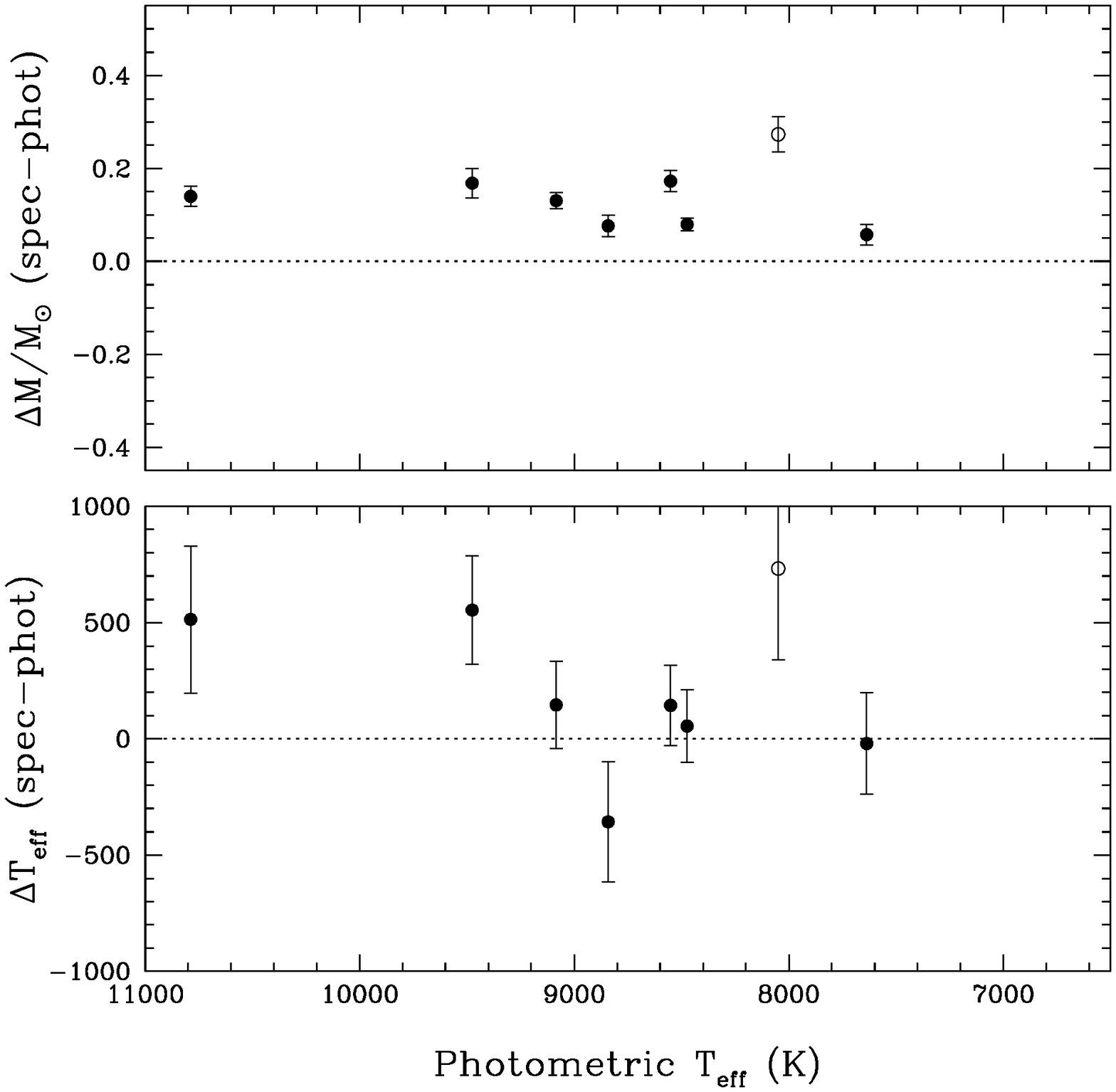] 
{Same as Figure \ref{fg:f11} but with a photometric sample drawn from
the survey of \citet{subasavage09} with $VRIJHK$ photometry and
accurate CCD trigonometric parallax measurements. The object
represented by a circle (LHS 4040) is discussed in the text.
\label{fg:f12}}

\figcaption[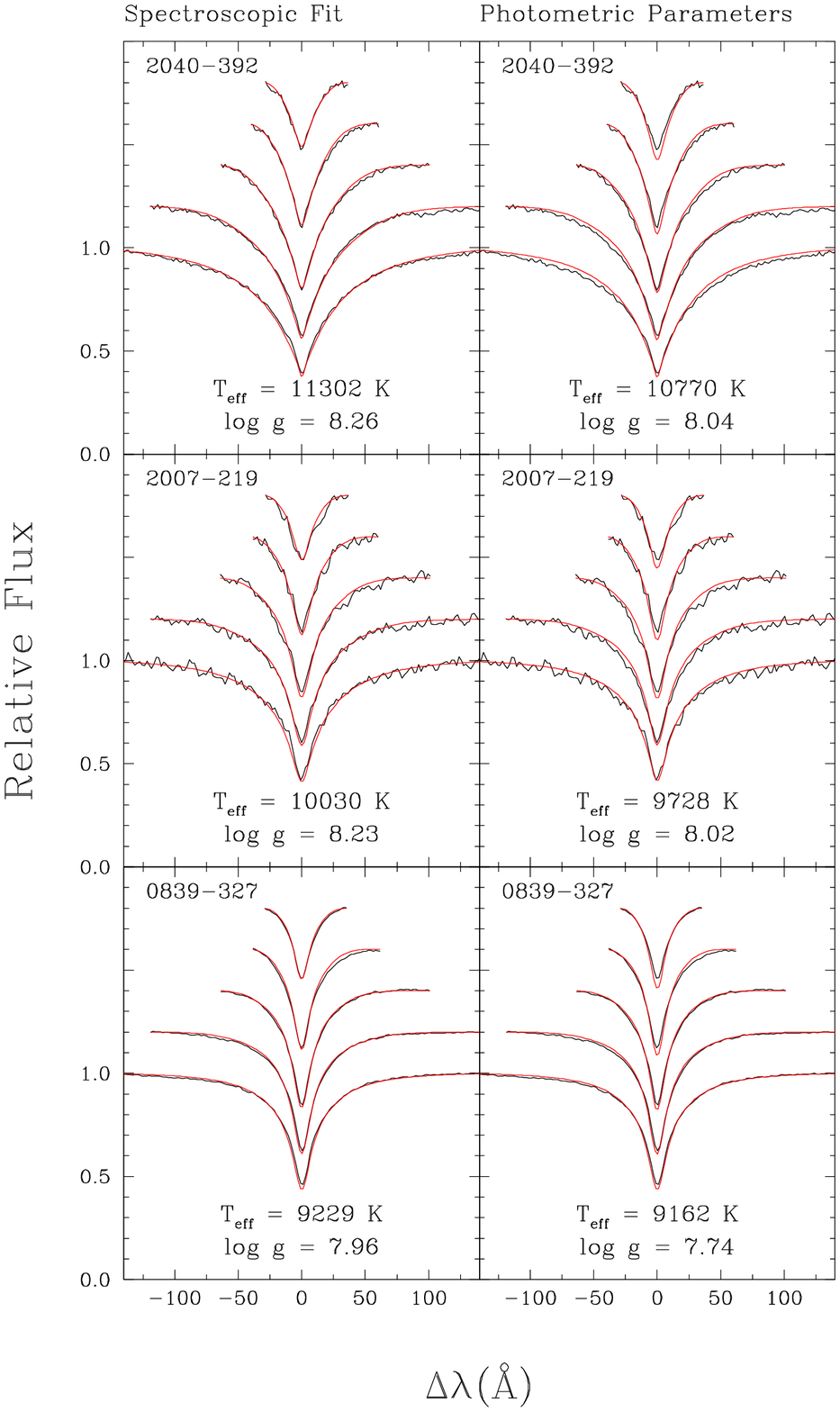] 
{Left panels: Best spectroscopic fits for three DA stars from the
sample of \citet{subasavage09}; the atmospheric parameters are given
in each panel. Right panels: Spectroscopic data for the same stars
compared with model spectra {\it interpolated} at the $\Te$ and
$\logg$ values obtained from the photometric method.
\label{fg:f13}}

\figcaption[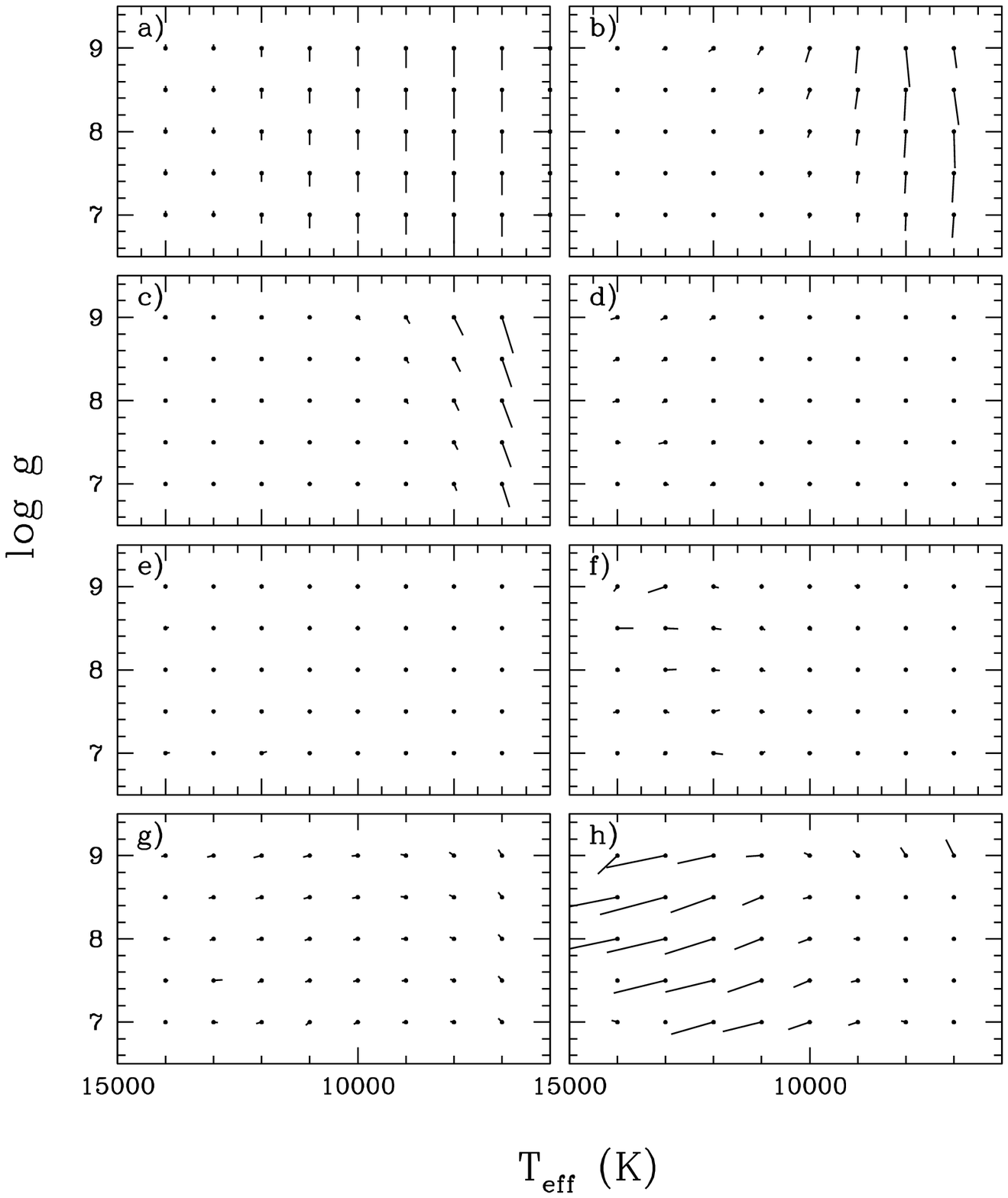] 
{Panel (a) illustrates the shift in spectroscopic $\logg$ values as a
function of $\Te$ required to have at each temperature a mean mass
comparable to that of hotter DA stars.  We assume for simplicity that
the shift is the same for all values of $\logg$ at a given
temperature. This panel serves as reference for the various
experiments described at length in the text, and whose results are
presented in the other panels.  In each case, our reference model grid
is fitted with a test grid, and the differences in the atmospheric
parameters are reported in the figure.
The changes in the models are: (b) the hydrogen radius taken as the
Bohr radius in the HM88 theory, (c) the neutral broadening enhanced by
a factor of two, (d) the \citet{potekhin02} proton microfield
distribution instead of the \citet{hooper68} distribution, (e) an
alternative cut-off for the non-ideal gas effects in the UV at 1215
\AA\ instead of 972 \AA, (f) the neglect of non-ideal effects due to 
electronic collisions in the Hummer-Mihalas formalism, (g) the H$^-$ 
opacity enhanced by 15\%, and (h) the ML2/$\alpha=1.5$ parameterization 
of the MLT.
\label{fg:f14}}

\begin{figure}[p]
\plotone{f1.eps}
\begin{flushright}
Figure \ref{fg:f1}
\end{flushright}
\end{figure}

\begin{figure}[p]
\plotone{f2.eps}
\begin{flushright}
Figure \ref{fg:f2}
\end{flushright}
\end{figure}

\begin{figure}[p]
\plotone{f3.eps}
\begin{flushright}
Figure \ref{fg:f3}
\end{flushright}
\end{figure}

\begin{figure}[p]
\plotone{f4.eps}
\begin{flushright}
Figure \ref{fg:f4}
\end{flushright}
\end{figure}

\begin{figure}[p]
\plotone{f5.eps}
\begin{flushright}
Figure \ref{fg:f5}
\end{flushright}
\end{figure}

\begin{figure}[p]
\plotone{f6.eps}
\begin{flushright}
Figure \ref{fg:f6}
\end{flushright}
\end{figure}

\begin{figure}[p]
\plotone{f7.eps}
\begin{flushright}
Figure \ref{fg:f7}
\end{flushright}
\end{figure}

\begin{figure}[p]
\plotone{f8.eps}
\begin{flushright}
Figure \ref{fg:f8}
\end{flushright}
\end{figure}

\begin{figure}[p]
\plotone{f9.eps}
\begin{flushright}
Figure \ref{fg:f9}
\end{flushright}
\end{figure}

\begin{figure}[p]
\plotone{f10.eps}
\begin{flushright}
Figure \ref{fg:f10}
\end{flushright}
\end{figure}

\begin{figure}[p]
\plotone{f11.eps}
\begin{flushright}
Figure \ref{fg:f11}
\end{flushright}
\end{figure}

\begin{figure}[p]
\plotone{f12.eps}
\begin{flushright}
Figure \ref{fg:f12}
\end{flushright}
\end{figure}

\begin{figure}[p]
\plotone{f13.eps}
\begin{flushright}
Figure \ref{fg:f13}
\end{flushright}
\end{figure}

\begin{figure}[p]
\plotone{f14.eps}
\begin{flushright}
Figure \ref{fg:f14}
\end{flushright}
\end{figure}

\end{document}